\documentclass[%
 reprint,
superscriptaddress,
 amsmath,amssymb,
 aps,pre,hyphens,floatfix
]{revtex4-1}

\usepackage[]{amsmath}
\usepackage{enumerate}
\usepackage[usenames,dvipsnames,svgnames,table]{xcolor}
\usepackage{IEEEtrantools}
\usepackage{graphicx}
\usepackage{array}
\usepackage{multirow}
\usepackage{verbatim}

\newcommand*{\tabbox}[2][t]{%
    \vspace{-8pt}\parbox[#1][12\baselineskip]{0.8cm}{\strut#2\strut}}

\begin{document}

\preprint{APS/123-QED}

\title{A coupled two-species model for the pair contact process with diffusion}

\author{Shengfeng Deng}
\email[]{gitsteven@gmail.com}
\affiliation{Key Laboratory of Quark and Lepton Physics (MOE) and Institute of Particle Physics, Central China Normal University, Wuhan 430079, China}
\affiliation{Department of Physics and Center for Soft Matter and Biological Physics, \\
	Virginia Tech, Blacksburg, VA 24061, USA}
\author{Wei Li}
\email[]{liw@mail.ccnu.edu.cn}
\affiliation{Key Laboratory of Quark and Lepton Physics (MOE) and Institute of Particle Physics, Central China Normal University, Wuhan 430079, China}
\author{Uwe C.~T\"{a}uber}
\email[]{tauber@vt.edu}
\affiliation{Department of Physics and Center for Soft Matter and Biological Physics, \\
	Virginia Tech, Blacksburg, VA 24061, USA}

\date{\today}

\begin{abstract}
The contact process with diffusion (PCPD) defined by the binary reactions 
$B + B \to B + B + B$, $B + B \to \emptyset$ and diffusive particle spreading 
exhibits an unusual active to absorbing phase transition whose universality class 
has long been disputed.
Multiple studies have indicated that an explicit account of particle pair degrees 
of freedom may be required to properly capture this system's effective long-time, 
large-scale behavior. 
We introduce a two-species representation for the PCPD in which single particles 
$B$ and particle pairs $A$ are dynamically coupled according to the stochastic 
reaction processes $B + B \to A$, $A \to A + B$, $A \to \emptyset$, and 
$A \to B + B$, with each particle type diffusing independently. 
Mean-field analysis reveals that the phase transition of this model is driven by 
competition and balance between the two species. 
We employ Monte Carlo simulations in one, two, and three dimensions to demonstrate 
that this model consistently captures the pertinent features of the PCPD.
In the inactive phase, $A$ particles rapidly go extinct, effectively leaving the 
$B$ species to undergo pure diffusion-limited pair annihilation kinetics 
$B + B \to \emptyset$.  
At criticality, both $A$ and $B$ densities decay with the same exponents (within 
numerical errors) as the corresponding order parameters of the original PCPD, and 
display mean-field scaling above the upper critical dimension $d_c = 2$. 
In one dimension, the critical exponents for the $B$ species obtained from seed 
simulations also agree well with previously reported exponent value ranges. 
We demonstrate that the scaling properties of consecutive particle pairs in the PCPD are
identical with that of the $A$ species in the coupled model.
This two-species picture resolves the conceptual difficulty for seed simulations in the 
original PCPD and naturally introduces multiple length and time scales to the system,
which are also the origin of strong corrections to scaling.
The extracted moment ratios from our simulations indicate that our model displays 
the same temporal crossover behavior as the PCPD, which further corroborates its full
dynamical equivalence with our coupled model.
\end{abstract}

\maketitle


\section{Introduction \label{sec1}}

The classification of phase transitions far from thermal equilibrium remains a 
very challenging problem that has attracted much interest
\cite{marro2005nonequi,odor2008universality,tauber2014critical}. 
In particular, nonequilibrium systems exhibiting phase transitions from active 
phases into inactive, absorbing states are both phenomenologically relevant, 
e.g., for species extinction in ecology, and fundamentally important for the 
understanding of a large variety of phenomena in nature
\cite{hinrichsen2000non,henkel2008non,henkel2011non}. 
Similar to equilibrium critical phenomena, only a few distinct universality 
classes have been established for active-to-absorbing-state phase transitions. 
The most prominent and generic universality class is that of critical directed 
percolation (DP) \cite{janssen1981noneqi,grassberger1982phase}; also 
well-established is the parity-conserving (PC) universality class 
\cite{cardy1998field}. 
The significance of active to absorbing phase transitions may be best 
exemplified by the DP universality class, the critical properties of which were 
perhaps most convincingly confirmed in electrohydrodynamic convection experiments
in liquid crystals \cite{takeuchi2007direc,takeuchi2009exp}. 
The DP class exhibits remarkable robustness with respect to detailed microscopic 
dynamical rules and has been widely discovered in very diverse phenomena, such 
as the extinction threshold for the Lotka-Volterra predator-prey competition 
model \cite{tauber2012pop,chen2016non}, the onset of turbulence in pipe flow 
\cite{shih2016ecologi,goldenfeld2017turbu}, the phase transition in glassy 
systems with blockages \cite{webman1998,vojta2012monte}, and even extends to 
systems with multiple species \cite{janssen2001dir}. 

Both the directed percolation and parity-conserving universality classes can be 
described in terms of diffusion-limited stochastic identical-particle reactions 
that include branching processes (e.g., $B \to B + B$ for DP and $B \to B + B + B$
for PC) and mutual pair annihilation ($B + B \to \emptyset$ for both DP and PC). 
The competition between these two types of reactions induces nonequilibrium phase
transitions from branching-dominated active states characterized by a fluctuating 
particle density with nonvanishing stationary average $\rho_s > 0$, to one or 
several absorbing states with $\rho_s = 0$, wherein annihilation processes prevail. 
A distinct universality class is to be expected when both branching and 
annihilation reactions require a \textit{pair} of particles to come into contact 
\cite{grassberger1982phase}. 
The \textit{pair contact process with diffusion} (PCPD) may be defined through the 
\textit{binary} reactions \cite{howard1997real}
\begin{equation}
	\textrm{fission: } B + B \to B + B + B , \ 
	\textrm{annihilation: } B + B \to \emptyset \, ,
	\label{eqs1}
\end{equation} 
in conjunction with individual particle diffusion, and some additional mechanism to 
prevent the particle density from diverging in the active state. 
If particles are not allowed to hop, this model reduces to the standard pair contact 
process which belongs to the DP universality class \cite{jensen1993critical}.

Both numerical simulations \cite{odor2002phase} and theoretical analysis 
\cite{howard1997real,janssen2004pair} suggest that the upper critical dimension of 
the PCPD is $d_c = 2$ and, in contrast to the exponential density decay in many 
other models, its inactive phase should be governed by the algebraic density decay 
of pure pair annihilation (for which $d_c = 2$ as well) 
\cite{howard1997real,janssen2004pair}:
\begin{equation}
	\rho(t) \sim
	\begin{cases}
		(D t)^{-1/2}       & d = 1 \\
		(D t)^{-1} \ln D t & d = d_c = 2 \\
		(\lambda t)^{-1}   & d > 2 \, , \\
    \end{cases}
\label{eqs2}
\end{equation}
with annihilation rate $\lambda$ and diffusivity $D$.

However, owing to extremely long crossover times and the ensuing notorious strong 
corrections to scaling \cite{odor2000critical,hinrichsen2001pair,
hinrichsen2006phase,kwon2007double,park2014critical}, the one-dimensional PCPD
has to date defied all attempts to conclusively unveil its critical properties (see 
Ref.~\cite{henkel2004non} for a comprehensive review of early studies). 
Consequently, numerical analyses have reported rather scattered values for the PCPD
critical exponents (c.f. corresponding tables in 
Refs.~\cite{henkel2004non,smallenburg2008univ}), and essentially three distinct
scenarios have been posited: 
(i) The PCPD constitutes a novel universality class
\cite{henkel2001univer,park2002well,odor2003criti,park2014critical} that exhibits 
robustness against nonuniversal modifications \cite{kockelkoren2003abs}; 
(ii) it belongs to the DP class 
\cite{hinrichsen2006phase,smallenburg2008univ,schram2012critical};
(iii) its critical characteristics may depend on the implemented diffusion rate 
\cite{odor2000critical,dickman2002nonun}.
To illustrate the severe predicaments in the PCPD numerical analysis, we note that 
according to Ref.~\cite{kwon2007double}, there are still significant corrections to 
scaling visible even after $10^{20}$ Monte Carlo steps.
Hence to perhaps ultimately clarify the asymptotic large-scale, long-time universal 
nature of the PCPD, field-theoretic methods may be indispensable, even imperative 
\cite{tauber2014critical}.
However, a direct implementation of the Doi--Peliti formalism to the competing 
single-species reactions \eqref{eqs1} and subsequent dynamical renormalization group 
analysis turned out unsuccessful, within both a perturbative approach 
\cite{janssen2004pair}, which yielded runaway renormalization group flow 
trajectories, as well as in a nonperturbative framework \cite{gredat2014finite}, 
which resulted in emergent nonanalyticity at a finite length scale. 
It is believed that these failures originate from the starting action, likely 
indicating the inadequacy of the single-species description for this system in 
constructing an appropriate coarse-grained action.

Hinrichsen pointed out that the PCPD stochastic processes effectively involve the 
interplay between two distinct species: one represents positively correlated 
particle pairs ($A$) assembling into fluctuating clusters, while the other 
represents solitary diffusive particles ($B$). 
Indeed, similar spatial-temporal patterns as observed in the original PCPD were 
found in a cyclically coupled two-species model \cite{hinrichsen2001cyc}. 
Another indication stems from the action of the driven PCPD
\cite{park2005driven,park2005cluster}, 
which, in stark contrast to driven variants of well-understood single-species models 
described by DP or PC, displays a violation of Galilean invariance.
Consequently, the external driving constitutes a relevant perturbation, resembling 
the situation in multispecies models. 
The manifest double domain structure \cite{kwon2007double} of the PCPD that 
demonstrates a strong correction to scaling for the dynamical exponent provides 
another perspective to justify the important role of particle pairs. 
Furthermore, the emergent linear coupling found in the nonperturbative 
renormalization group approach due to the finite-scale singularity may be 
related to ``elementary excitations'' consisting of bound particle pairs that would 
involve a separate intrinsic scale \cite{gredat2014finite}. 

Therefore it appears desirable to construct a more \textit{fine-grained two-species}
model representing the PCPD, which properly accounts for its internal stochastic 
noise generated by the reactions.
Any asymptotic (perhaps hidden) symmetry \cite{henkel2004non} of the PCPD might 
then be revealed by further coarse-graining to interacting continuous fields. 
We remark that this situation is reminiscent of phase transitions in antiferromagnets, 
where prior to coarse-graining the microscopic Heisenberg spin Hamiltonian and 
taking the continuum limit, the staggered magnetization must first be introduced as 
the proper order parameter, which then becomes dynamically coupled to the 
conserved magnetization (model G \cite{folk2006critical,tauber2014critical}). 
As noted above, previously a cyclically coupled model was proposed by Hinrichsen 
\cite{hinrichsen2001cyc}, in which the subsystem of pair particles follows a DP 
process and moreover, pairs $A$ are allowed to transmute into solitary particles $B$ 
through the reaction $A \to B$. 
Yet these peculiar model features render its connections with both DP and PCPD vague 
\cite{hinrichsen2003stoc}. 
Several other multispecies extensions of the PCPD were studied in 
Ref.~\cite{odor2002multi}; though one of these extensions was reported to behave
like the PCPD, it did not entail a separation of particle pairs and solitary particles.

In this paper, we introduce a more transparent two-species model that can be more 
straightforwardly linked to the original PCPD.
To this end, by introducing a species $A$ that represents pairs of particles, and 
identifying pair production as an intermediate process $B + B \to A$, the PCPD 
reactions in \eqref{eqs1} can be reinterpreted in terms of the stochastic processes 
$A \to B + B + B$ and $A \to \emptyset$. 
Naturally, the pair splitting process $A \to B + B$ should be allowed to happen as 
well, whence the reaction $A \to B + B + B$ may effectively be replaced by 
$A \to A + B$ (or alternatively, $A \to A + A$, since parity conservation is 
irrelevant here \cite{park2001binary}). 
Consequently, we henceforth study the following \textit{coupled two-species model} 
(CPCPD) by means of both mean-field analysis and Monte Carlo simulations:
\begin{equation}
	B + B \stackrel{\tau}{\to} A \, , \quad  
	A \stackrel{\sigma}{\to} A + B \, , \quad
	A \stackrel{\mu}{\to} \emptyset \, , \quad 
	A \stackrel{\rho}{\to} B + B \, .
\label{eqs3}
\end{equation}
In order to restrict the total particle density, preventing its divergence in the 
active phase, one may resort either to hardcore site exclusion for the particles as 
is implemented in most numerical simulations, and in this present work (see 
Refs.~\cite{janssen2004pair,van2001field} for the corresponding field-theoretic 
treatment); or alternatively to soft restrictions, e.g., through higher-order 
annihilation reactions \cite{kockelkoren2003abs,tauber2012pop}. 

It is apparent that the above reaction scheme \eqref{eqs3}, accompanied by proper 
density restrictions, satisfies the general features of the PCPD 
\cite{hinrichsen2001cyc}:
(i) solitary particles $B$ are purely diffusive; 
(ii) reproduction requires a pair of neighboring particles; 
(iii) particles are removed if at least two particles come to contact at neighboring
sites; and 
(iv) there is a mechanism to limit the total particle density in the active phase.
For our CPCPD model, conditions (ii) and (iii) are met through the intermediate pair 
productions. 
Thus the CPCPD should be expected to display the same critical properties as the 
original PCPD and at the same time retain all its hallmarks, such as strong 
corrections to scaling; the interplay between two different dynamical modes and the 
manifestation of a double domain structure; pure pair annihilation kinetics in the 
inactive phase; etc., which will be examined in the following sections. 
Crucially, the two-species picture of the CPCPD also endows it with additional
insight and more natural interpretations for the peculiar properties of the original 
PCPD.
We shall not attempt here to posit conclusive assertions about the ensuing universal 
properties of either the PCPD or the CPCPD, but will rather present convincing 
evidence that both these models are consistent with each other in all studied 
aspects. 
We thereby hope to provide a new avenue to stimulate further research. 

The remainder of this paper is organized as follows:
In Sec.~\ref{sec2} we analyze the mean-field behavior of the CPCPD model. 
In Sec.~\ref{sec3} we detail our simulation method and compute various scaling 
exponents both at criticality and in the inactive phase. 
We also present numerical results for the temporal evolution of certain moment ratios 
\cite{dickman1998moment}, employing the quasi-stationary method
\cite{de2005simulate}. 
These exponent and moment ratio results are then compared with the corresponding 
known results for the original PCPD model. 
Finally, Sec.~\ref{sec4} summarizes this work and provides a brief outlook.

\section{Mean-field analysis \label{sec2}}

Denoting the number of $A$ and $B$ particles at lattice site $i$ at time $t$ as 
$a_i(t)$ and $b_i(t)$, where the index $i$ could represent a $d$-dimensional position
label, we consider as order parameters the mean densities of $A$ and $B$ particles,
\begin{equation}
	a(t) = \left\langle \frac{1}{L^d} \sum_{i} a_i(t) \right\rangle , \quad
	b(t) = \left\langle \frac{1}{L^d} \sum_{i} b_i(t) \right\rangle ,
	\label{eqs:dens}
\end{equation}
where $L$ represents the linear system size, and the brackets $\langle\cdots\rangle$
indicate ensemble averages. 
Within a mean-field framework, site exclusions are not necessarily implemented as
they usually are in simulations. 
One may instead invoke soft restrictions \cite{tauber2012pop} to limit the particle 
densities in the active phase; the precise microscopic scheme to implement these 
particle density restrictions should not affect large-scale universal properties. 
To demonstrate this, we consider the following three distinct versions of triplet 
annihilation reactions for the original PCPD model, 
$B + B + B \to \emptyset / B / B + B$, which in our effective two-species CPCPD 
model translate to
\begin{equation}
	A + B \xrightarrow{\nu/\nu'/\tilde{\nu}} \emptyset / B / A \, ,
\label{eqs5}
\end{equation}
with reaction rates $\nu$, $\nu'$, and $\tilde{\nu}$, respectively. 
For the reaction scheme \eqref{eqs3} and \eqref{eqs5}, the corresponding mean-field 
rate equations for the average densities $a(t)$ and $b(t)$ read
\begin{IEEEeqnarray}{rCl}
	\label{eqs6}
        \frac{\partial a(t)}{\partial t} &=& - \left( \mu + \rho \right)
		a(t) + \tau b(t)^2 - \left( \nu + \nu' \right) a(t) b(t), \qquad
	\IEEEyesnumber* \IEEEyessubnumber*
	\label{eqs6a}\\
        \frac{\partial b(t)}{\partial t} &=& 
        \left( \sigma + 2\rho \right) a(t) - 2\tau b(t)^2 
		- \left( \nu + \tilde{\nu} \right) a(t) b(t) \, . \qquad
	\label{eqs6b}
\end{IEEEeqnarray}

\subsection{Order parameter and density decay exponents \label{sec2.1}}

The mean-field rate equations \eqref{eqs6} yield two stationary solutions:
\begin{IEEEeqnarray}{rCl}
	\label{eqs7}
	b_{s} &=& 0 \, , \quad a_{s} = 0 \, ; 
	\IEEEyesnumber* \IEEEyessubnumber*
	\label{eqs7a} \\
	b_{s} &=& \frac{\sigma-2\mu}{3\nu+2\nu'+\tilde{\nu}} \, , \nonumber \\
	a_{s} &=& \frac{\tau(\sigma-2\mu)^2/(3\nu+2\nu'+\tilde{\nu})}
		{(3\nu+2\nu'+\tilde{\nu})\rho
		+(\nu+\tilde{\nu})\mu+(\nu+\nu')\sigma} \, . \qquad 
	\label{eqs7b}
\end{IEEEeqnarray}
The solution \eqref{eqs7a} of course represents the inactive, absorbing state, stable
for $\sigma < 2\mu$, and includes the critical stationary state at $\sigma = 2\mu$. 
The stationary solution \eqref{eqs7b}, on the other hand, corresponds to the active 
phase which requires $\sigma > 2\mu$. 
Note that the critical point is determined by the balance of the branching and 
annihilation processes, and is hence located at $\sigma_c = 2\mu$. 
However, the mechanism that drives the (C)PCPD system into the critical state needs 
to be distinguished from that of DP and other single-species models, where the 
balance of these processes directly manifests itself in the vanishing of a linear 
(``mass'') term in the rate equation. Consequently, a critical algebraic power 
law decay sets in. In an unrestricted ``bosonic'' representation of the PCPD 
\cite{howard1997real}, the competing binary particle production and annihilation 
reactions balance precisely at the critical point, and lead to a constant stationary 
density there. In the CPCPD, instead the critical point is induced by a precise match 
between the density changes of both distinct species: $-2 \Delta a = \Delta b$.
Upon neglecting the higher-order terms originating from the soft density 
restrictions, this balance leads to 
\begin{equation*}
    -2 \underbrace{\big[-(\mu+\rho)a+\tau b^2\big]\Delta t }_{\Delta a} =
    \underbrace{\big[(\sigma_c+2\rho)a-2\tau b^2\big]\Delta t}_{\Delta b} \, , 
\end{equation*}
whence indeed $\sigma_c = 2\mu$.
In the active phase near the transition point, the stationary densities scale as
\begin{equation}
	a_s \sim \Delta^{\beta_a} \, , \qquad b_s \sim \Delta^{\beta_b} \, ,
	\label{eqs8}
\end{equation}
where $\Delta = \sigma - \sigma_c$.
From Eq.~\eqref{eqs7b}, we immediately infer $\beta_a = 2$ and $\beta_b = 1$.

\begin{figure}[!tbp]
	\begin{center}
		\includegraphics[width=0.45\textwidth]{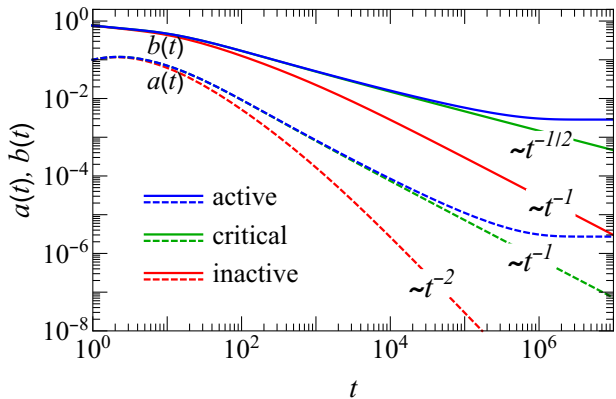}
	\end{center}
	\caption{Density decay curves for the mean pair density $a(t)$ (dasded curves) 
	and the mean single-particle density $b(t)$ (solid) in the active phase (blue 
	curves with $\sigma = 0.602$), at the critical point (green curves with 
	$\sigma = 0.6$), and in the inactive phase (red curves with $\sigma = 0.5$); 
	with the other rates held fixed at $\mu = 0.3$, $\tau = 0.2$, $\rho = 0.3$, 
	$\nu = 0.1$, $\nu' = 0.1$, $\tilde{\nu} = 0.2$, and initial densities $a(0) = 0$
	and $b(0) = 1$.}
	\label{fig:decay}
\end{figure}
In Fig.~\ref{fig:decay} we show the numerical solutions of the coupled rate equations
\eqref{eqs6}. 
In the active phase (blue curves), the densities exponentially relax to the 
stationary nonzero values \eqref{eqs7b}, as also can be checked by a linear expansion
of Eqs.~\eqref{eqs6} around the stationary values \eqref{eqs7b}.
In the asymptotic long-time regime, both $a(t)$ and $b(t)$ decay algebraically in the
entire inactive phase (red curves),
\begin{equation}
	a(t) \sim t^{-\delta_{\mathrm{in}a}} \, , \qquad b(t) \sim 
	t^{-\delta_{\mathrm{in}b}} \, ,
	\label{eqs9}
\end{equation}
as well as at the critical point (green lines), with decay exponents 
$\delta_a$ and $\delta_b$.
In the inactive phase, one obtains $\delta_{\mathrm{in}a} = 2$ and
$\delta_{\mathrm{in}b} = 1$ for the pair and individual particle densities $a(t)$ and
$b(t)$, respectively. 
The decay exponent $\delta_{\mathrm{in}b} = 1$ recovers precisely the mean-field 
PCPD value in the inactive phase for $d > 2$,~c.f.~Eqs.~\eqref{eqs2}: 
$B$ particles undergo pure pair annihilation processes, while the $A$ species dies 
out more quickly since $\delta_{\mathrm{in}a} = 2 \delta_{\mathrm{in}b}$. 
At the critical point $\sigma_c = 2\mu$, the decay exponent for the single $B$ 
particles, $\delta_b = 1/2$, is again identical to the corresponding value for the 
mean-field PCPD \cite{janssen2004pair,odor2002phase}, but $\delta_a = 1$ differs 
from the mean-field parity-conserving PCPD value measured in $d_c = 2$ 
dimensions, which was reported to be identical with $\delta_b$ \cite{odor2002phase}. 
These mean-field decay patterns for $a(t)$ and $b(t)$ both in the inactive phase and
at the critical point are also observed in simulations in the two- and 
three-dimensional CPCPD (as detailed in the following sections), with corresponding 
logarithmic corrections to the mean-field scaling at $d_c = 2$. 
The local slopes of the mean-field decay curves are shown in App.~\ref{appda};
they indicate that the slow crossover behavior of the (C)PCPD even persists in the 
mean-field regime.
Yet in one dimension, fluctuations become strongly enhanced and invalidate the above
mean-field results. 
In the inactive phase, $\delta_{\mathrm{in}b}$ is then expected to be $1/2$ as in 
Eq.~\eqref{eqs2}.
At the critical point, one would anticipate $\delta_a = \delta_b$ as 
demonstrated in various simulation studies, c.f., for example, 
Ref.~\cite{hinrichsen2001pair}; this is demonstrated in Figs.~\ref{fig:crit}(a) 
and (b) below as well.

We remark that the coupled nonlinear first-order ordinary differential equations 
\eqref{eqs6} can in principle be solved analytically (see the discussion in 
Ref.~\cite{chandrasekar2009comp}). 
The form of these solutions will of course split into distinct scenarios depending on
which phase one studies. 
The asymptotic scaling properties may then be extracted from the leading-order terms. 
Yet pursuing that complete solution would greatly divert from the main object of this
paper; we therefore only show that, in the inactive phase, one can approximately solve
Eqs.~\eqref{eqs6} and subsequently obtain the decay exponents 
$\delta_{\mathrm{in}a} = 2$ and $\delta_{\mathrm{in}b} = 1$. 
To this end, notice that the first term of Eq.~\eqref{eqs6b} can be regarded negligible
as compared to the first term in Eq.~\eqref{eqs6a}. 
Furthermore, we may also neglect the higher-order terms. 
This leads to the following approximate differential equations in the inactive phase: 
\begin{IEEEeqnarray}{rCl}
	\label{eqs10}
        \frac{\partial a(t)}{\partial t} &\approx& 
		- \left( \mu+\rho \right) a(t) + \tau b(t)^2 \, , 
	\IEEEyesnumber* \IEEEyessubnumber*
	\label{eqs10a} \\
        \frac{\partial b(t)}{\partial t} &\approx& - 2 \tau b(t)^2 \, .
	\label{eqs10b}
\end{IEEEeqnarray}
With initial conditions $a(0) = a_0$ and $b(0) = b_0$, these are solved by
\begin{IEEEeqnarray}{rCl}
	\label{eqs11}
	b(t) &=& \frac{b_0}{1+2 b_0 \tau t} \sim t^{-1} 
	\IEEEyesnumber* \IEEEyessubnumber* \, , \\
	a(t)&=& \frac{1}{4 \tau (1 + 2 b_0 \tau t)}
	e^{-\frac{(\mu +\rho)(1+2 b_0 \tau t)}{2 b_0 \tau}} 
	\Bigg[ 2 \tau  e^{\frac{\mu+\rho }{2 b_0 \tau}} \nonumber \\
	& & \times\left[ \left( 2 a_0+b_0 \right) \left( 1 + 2 b_0 \tau t \right) 
	- b_0 e^{(\mu +\rho )t} \right] \nonumber \\
	& & + (\mu +\rho) \left( 1+2 b_0 \tau t \right) 
	\Bigg( \mathrm{Ei} \left[ \frac{1}{2} (\mu +\rho) 
	\left( 2 t+\frac{1}{b_0 \tau} \right) \right] \nonumber \\ 
	& & - \mathrm{Ei}\left[ \frac{\mu +\rho }{2 b_0 \tau} \right] \Bigg)
	\Bigg] \sim t^{-2} \, ,
\end{IEEEeqnarray}
where the exponential integral function is defined as
$\mathrm{Ei}(x) = -\int_{-x}^{\infty} \frac{e^{-z}}{z} \, dz$.

\subsection{Scaling analysis \label{sec2.2}}

We next extend the mean-field analysis to spatially extended systems.
To this end, we consider local reactions, and coarse-grained particle densities $a(x,t)$
and $b(x,t)$, with diffusive particle propagation. 
The coupled rate equations \eqref{eqs6} are then generalized to the following 
mean-field reaction-diffusion partial differential equations:
\begin{IEEEeqnarray}{rCl}
	\label{eqs12}
        \frac{\partial a(x,t)}{\partial t} &=& D_A \nabla^2 a(x,t) 
		- \left( \mu + \rho \right) a(x,t) \nonumber \\
		& & + \tau b(x,t)^2
		- \left( \nu + \nu' \right) a(x,t) b(x,t) \, , \qquad
		\IEEEyesnumber* \IEEEyessubnumber* \\
        \frac{\partial b(x,t)}{\partial t} &=& D_B \nabla^2 b(x,t) 
		+ \left( \sigma + 2\rho \right) a(x,t) \nonumber \\
		& & - 2 \tau b(x,t)^2 
		- \left( \nu + \tilde{\nu} \right) a(x,t) b(x,t) \, . \qquad
\end{IEEEeqnarray}
Here the particle densities spread with diffusion rates $D_A$ and $D_B$, 
respectively.

At the critical point, the scale-invariant nature of the system renders 
Eqs.~\eqref{eqs12} invariant under the rescaling of the coordinates 
$x \to \Lambda x$, combined with appropriate rescalings of time, the densities, and 
the reaction rates \cite{hinrichsen2000non}. 
To obtain these rescaling relations, we first note that upon approaching the critical
point, the correlation lengths $\xi_{a}$ and $\xi_{b}$, and the characteristic time 
scales $t_{c a}$ and $t_{c b}$, diverge as
\begin{IEEEeqnarray}{rCl}
	\xi_{a} \sim \Delta^{-\nu_{a}} \, &,& \qquad 
	\xi_{b} \sim \Delta^{-\nu_{b}} \, , 
	\IEEEyesnumber* \IEEEyessubnumber* \label{eqs:corrs} \\
   	t_{c a} \sim \Delta^{-z_a \nu_a} \, &,& \qquad
	t_{c b} \sim \Delta^{-z_b \nu_b} \, .
	\label{eqs:corrt}
\end{IEEEeqnarray}
We have presumed that there are in general two sets of scaling exponents pertaining 
to each species.
Within the mean-field approximation, diffusive spreading implies 
\[ z_a = z_b = 2 \, . \]

Then, with the near-critical density scaling \eqref{eqs8} and the correlation length
scaling \eqref{eqs:corrs}, we infer the following scaling behavior:
\begin{eqnarray*}
	x \to \Lambda x \, , \quad t \to \Lambda^2 t \, , \quad 
	a \to \Lambda^{-\beta_a/\nu_{a}} a \, , \\
	b \to \Lambda^{-\beta_b/\nu_{b}} b \, , \quad
	\mu \to \Lambda^{-2} \mu \, , \quad 
	\sigma \to \Lambda^{-2} \sigma \, , \\ 
	\rho \to \Lambda^{-2} \rho \, , \quad 
	\tau \to \Lambda^{-x_\tau} \tau \, , \quad 
	\nu \to \Lambda^{-x_\nu} \nu \, .
\end{eqnarray*}
Demanding scale invariance for Eqs.~\eqref{eqs12} thus yields the relations
\begin{IEEEeqnarray*}{rCl}
	x_\tau + 2\beta_b/\nu_{b} &=& 2 + \beta_a/\nu_{a} \, , 
	\quad x_{\nu} + \beta_b/\nu_{b} = 2 \, , \\
	x_{\tau} + \beta_{b}/\nu_{b} &=& 2 \, , \quad
	x_{\nu} + \beta_a/\nu_{a} = 2 \, .
\end{IEEEeqnarray*}
Consistency requires $x_\tau = 0 = x_\nu$ and 
$\beta_a/\nu_{a} = 2 = \beta_{b}/\nu_{b}$. 
Inserting $\beta_a = 2$ and $\beta_b = 1$ from Eqs.~\eqref{eqs8}, we obtain
\begin{IEEEeqnarray}{rCl}
	\nu_{a} &=& 1 \, , \quad \nu_{b} = 1/2 \, .
	\label{eqs:nu}
\end{IEEEeqnarray}
These exponents are to be compared to those of the standard PCPD in the mean-field 
approximation, for which $\delta = 1/2$ ($= \delta_b$), $\beta = 1$ ($= \beta_b$), 
$z = 2$, and $\nu_{\perp} = 1$ ($ = \nu_a$) for identical particles 
\cite{janssen2004pair}; and to those obtained from cluster mean-field approximations,
which gave $\delta_1 = 1/2$ ($= \delta_b$), $\delta_2 = 1$ ($= \delta_a$), 
$\beta_1 = 1$ ($= \beta_b$) and $\beta_2 = 2$ ($= \beta_a$) 
\cite{odor2002phase}, where the indices ``1'' and ``2'' denote the exponents for 
single-particles and consecutive pairs of the PCPD respectively; see also 
Eq.~\eqref{eqs:rho}. 
It should be noted that the critical exponents $\nu_{a,b}$ and 
$\beta_{1,2,a,b}$ only pertain to approaching the transition from the active side. 
There is no finite correlation lengths in the inactive phase. 
In the more ``coarse-grained'' single-particle representation (the original PCPD), one
only sees the larger correlation length, hence $\nu = 1$ ($= \nu_a$). 
Our two-species picture for the CPCPD in contrast leads to multiple length and time 
scales corresponding to various correlations and cross-correlations between the two 
species of particles. 
When single values for $z$, $\nu$, etc., are measured in standard PCPD simulations, 
automatically the largest length and time scales, given by the longest correlation
length and characteristic relaxation time in the system, are singled out, masking other 
faster processes with shorter characteristic length and time scales.
In the CPCPD, these scales are set by $\xi_a$ and $t_{c a}$, since the particle 
\textit{pairs} $A$ represent the critical degrees of freedom that are subject to 
diverging correlations and critical slowing-down.
In Sec.~\ref{sec3.3}, we shall explore the implication of the presence of additional 
length and time scales in more detail.

\section{Numerical simulations \label{sec3}}

\subsection{Simulation method \label{sec3.1}}

To quantitatively study the competing stochastic processes \eqref{eqs3}, we have
conducted Monte Carlo simulations on one-, two- and three-dimensional (cubic) 
lattices with periodic boundary conditions, where each lattice site may either be 
empty or occupied by at most one particle of either the $A$ or $B$ species: 
$a_i(t)=0,1$ and $b_i(t)=0,1$. 
In order to reduce the overhead of generating random numbers, we modify the 
reactions \eqref{eqs3} by generalizing the simulation scheme of  
Ref.~\cite{carlon2001crit} for the original PCPD model. 
For example, in one dimension, the states of the lattice sites are updated in 
a random sequential manner according to these rules:
\begin{widetext}
\begin{IEEEeqnarray}{lll}
\label{eqs15}
 A\emptyset \leftrightarrow \emptyset A , \ B \emptyset
 \leftrightarrow \emptyset B  \quad &\textrm{ with rates }\ & 
 D_A \textrm{ and } D_B \, , 
 \IEEEyesnumber* \IEEEyessubnumber* \label{eqs15a}\\
 A\to \emptyset & \textrm{ with rate } & \mu(1-D_A-D_B) \, , \label{eqs15b}\\
 A \emptyset , \ \emptyset A \to AB , \ BA & \textrm{ with rate } &
 \sigma(1-\mu) (1-D_A-D_B)/2 \, , \label{eqs15c}\\
 BB\to A \emptyset , \ \emptyset A & \textrm{ with rate } &
 \tau(1-\sigma) (1-\mu) (1-D_A-D_B)/2 \, , \label{eqs15d}\\
 A\emptyset , \ \emptyset A \to BB & \textrm{ with rate } &
 (1-\tau) (1-\sigma) (1-\mu) (1-D_A-D_B)/2 \, , \qquad \label{eqs15e}
\end{IEEEeqnarray}
\end{widetext}
where the parameters $D_A$, $D_B$, $\sigma$, and $\tau$ are held at 
positive fixed values. 
One then only needs to tune the parameter $\mu$ to obtain different phases of 
the system.

It should be noted that the parameters are constrained by $0<D_A+D_B<1$, 
$0<\mu<1$, $0<\sigma<1$, and $0<\tau<1$. 
These rules altogether comprise $11$ reactions whose respective probabilities 
add up to unity, and are utilized to divide the uniform random number space 
$(0, 1)$ into $11$ sections. 
Except for the annihilation reaction \eqref{eqs15b}, a direction is also decided 
along with the reaction selection, so that the diffusion \eqref{eqs15a} / branching 
\eqref{eqs15c} / coagulation \eqref{eqs15d} / splitting \eqref{eqs15e} processes 
take place along this direction, provided that the corresponding reaction is allowed
by the configuration of the selected site and its neighbor in the chosen direction. 
In each update, we first randomly pick a site in the lattice, and then draw another
random number to decide which subsequent reaction may take place. 
Any process is aborted if it is not permissible and a new attempt starts. 
The simulation time is increased by $1$ after each Monte Carlo sweep for $L$ 
attempts. 
Simulation schemes in two and three dimensions can be constructed in a similar
fashion.

\begin{figure}[!tbp]
	\begin{center}
		\includegraphics[width=0.45\textwidth]{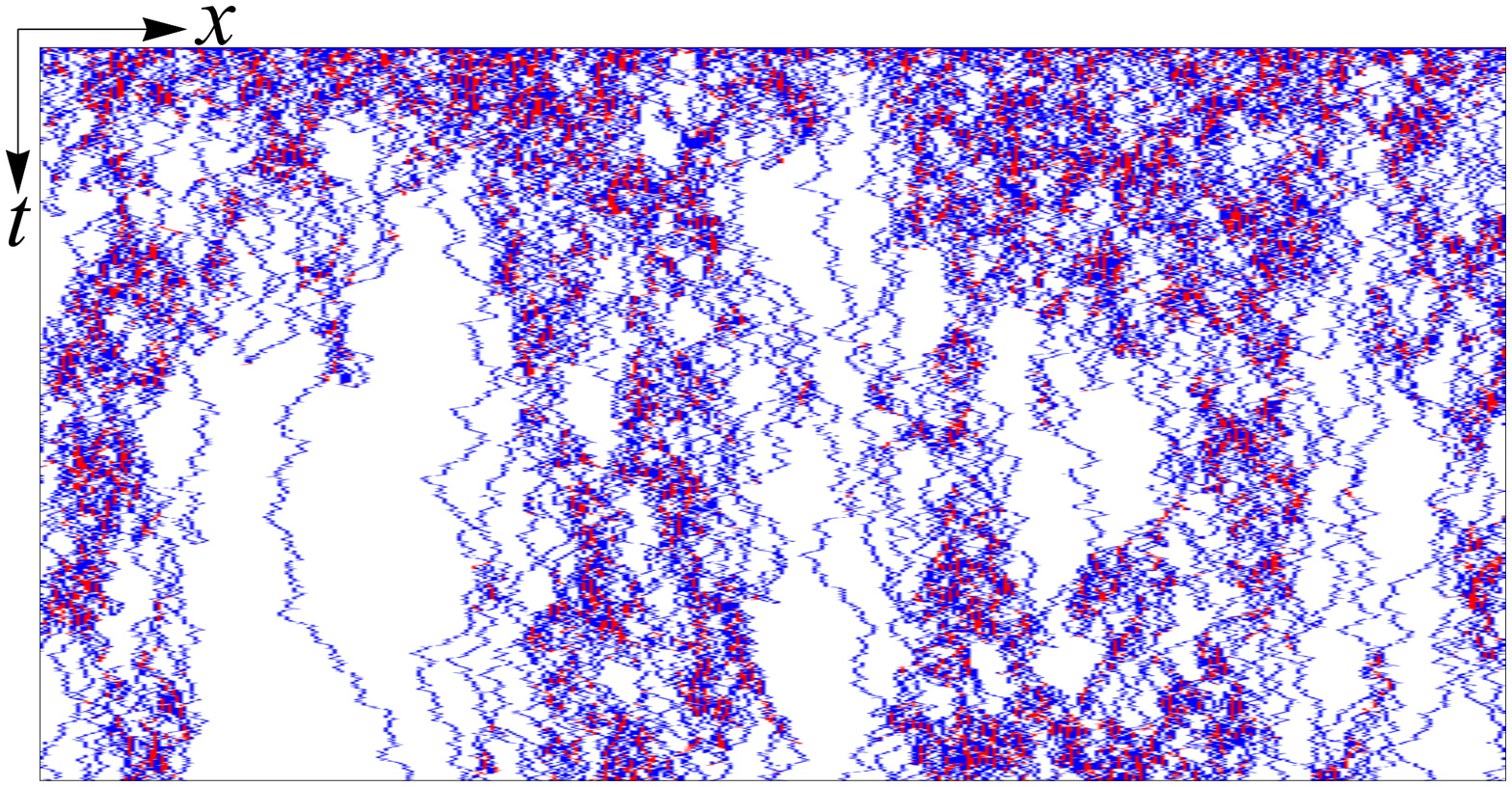}
	\end{center}
	\caption{A critical CPCPD process in one spatial dimension (horizontal) evolving
       in time (plotted vertically downward) starting from a lattice fully occupied by 
       $B$ particles. 
 	The $A$ and $B$ species are colored in red and blue, respectively.}
	\label{fig:2}
\end{figure}
Figure \ref{fig:2} shows a typical critical CPCPD process in one dimension starting 
from a lattice fully occupied by $B$ particles. 
This spatio-temporal pattern, which features intriguing interplays between solitarily
diffusing particles $B$ and highly active clusters consisting of pairs $A$ intensely
interacting with single-particle species $B$, visually closely resembles the original 
PCPD system. 

In order to compare our two-species CPCPD with the original PCPD model, we have 
also simulated the PCPD in one, two, and three dimensions, following 
Ref.~\cite{carlon2001crit}. 
The one-dimensional PCPD process, for example, is implemented with the dynamics
\begin{IEEEeqnarray}{lCl}
	\label{eqs16}
	B\emptyset\leftrightarrow \emptyset B \ &\textrm{ with rate } & D\, , 	
	\IEEEyesnumber* \IEEEyessubnumber* \\
	BB\to\emptyset\emptyset \ &\textrm{ with rate } & \mu (1-D)\, ,
	\label{eqs16b}\\
	\!\!\!\!\!\!\!\!\!\!
	BB\emptyset, \emptyset BB\to BBB \ &\textrm{ with rate } &
	(1-\mu) (1-D)/2\, , \ \label{eqs16c}
\end{IEEEeqnarray}
where $0<D<1$ and $0<\mu<1$. Again, upon selection of a reaction, a
direction is also decided and the above scheme altogether includes $6$ 
reactions. This implementation is effectively equivalent to the
algorithm in Ref.~\cite{odor2003criti} where a particle and a direction
are randomly selected first before a reaction is performed.
It is straightforward to extend the above scheme to simulations in 
two and three dimensions. 

For the particle production process, one may generate the new particle(s)
either along the chosen direction as the above scheme or evenly at the available 
neighboring sites. 
For simplicity, we implemented the former realization. 
In App.~\ref{appdb} we show results obtained with both algorithm variants for the 
parity-conserving PCPD; the critical properties are insensitive to these modifications.

\subsection{Density decay exponents \label{sec3.2}}

The PCPD displays nontrivial scaling features both in the inactive phase and at
criticality.
We begin with a comparison of the density decay exponents of the CPCPD and the
PCPD.
To this end, we follow the recent elaborate study \cite{park2014critical} and adopt the
notion that the PCPD represents a unique universality class, irrespective of the 
diffusivity \footnote{Upon inspecting the previously reported critical exponents
\cite{henkel2004non,smallenburg2008univ}, one notes that some early studies already
obtained consistent values for the critical decay exponent ($\alpha$ or $\delta$ in the
literature) and the critical order parameter growth exponent $\beta$ for the stationary
particle density \cite{odor2003criti}. 
In contrast, the critical exponents that entail the correlation length exponent 
$\nu_{\perp}$, such as $z = \nu_{\parallel}/\nu_{\perp}$ and $\beta/\nu_{\perp}$,
appeared to show remarkable dependences on the diffusion rate. 
Once we have provided numerical evidence that establishes the equivalence between 
the CPCPD and the PCPD, we will explore possible causes for this subtle distinction in
more detail in, c.f.~Sec.~\ref{sec3.3}.}.
Hence for all the Monte Carlo simulations reported in the following, we chose a
sufficiently large hopping probability $D = 0.7$ for the PCPD, and the also somewhat
arbitrarily selected diffusivities $D_A = 0.2$ and $D_B = 0.22$ for the CPCPD in one,
two, and three dimensions, unless otherwise explicitly specified. 
For the CPCPD we further set the reaction probabilities at fixed values $\sigma = 0.6$
and $\tau = 0.5$, whence $\mu$ becomes the sole control parameter for both the
CPCPD and the PCPD: the inactive, absorbing state appears for large $\mu$, while the
active phase is observed for small $\mu$. 
It is well established that considerable numerical effort is required to accurately locate
the critical point as well as to obtain reliable and consistent estimates for the PCPD 
critical exponents \cite{hinrichsen2003stoc,hinrichsen2006phase,park2014critical}.
In this study, we resort to comparing the CPCPD and PCPD scaling exponents with
moderate accuracy. 
We conducted our simulations with reasonably large linear system sizes $L$ and
Monte Carlo run times:
In one dimension, we employed $L=10,000$ and observables were measured up to
$t_{\mathrm{max}}=10^6$ Monte Carlo sweeps (MCS); in two dimensions, we used
the system length $L = 640$ and $t_{\mathrm{max}}=10^5$ MCS; in three
dimensions, we took $L = 320$ and also $t_{\mathrm{max}}=10^5$ MCS. 

\begin{figure*}[t]
   \begin{center}
	   \begin{tabular}{p{0.08\textwidth}p{0.44\textwidth}p{0.44\textwidth}}
  & \multicolumn{1}{c}{CPCPD} & \multicolumn{1}{c}{PCPD} \\
  \tabbox[c]{$d=1$} & \tabbox{\includegraphics[width=0.4\textwidth]{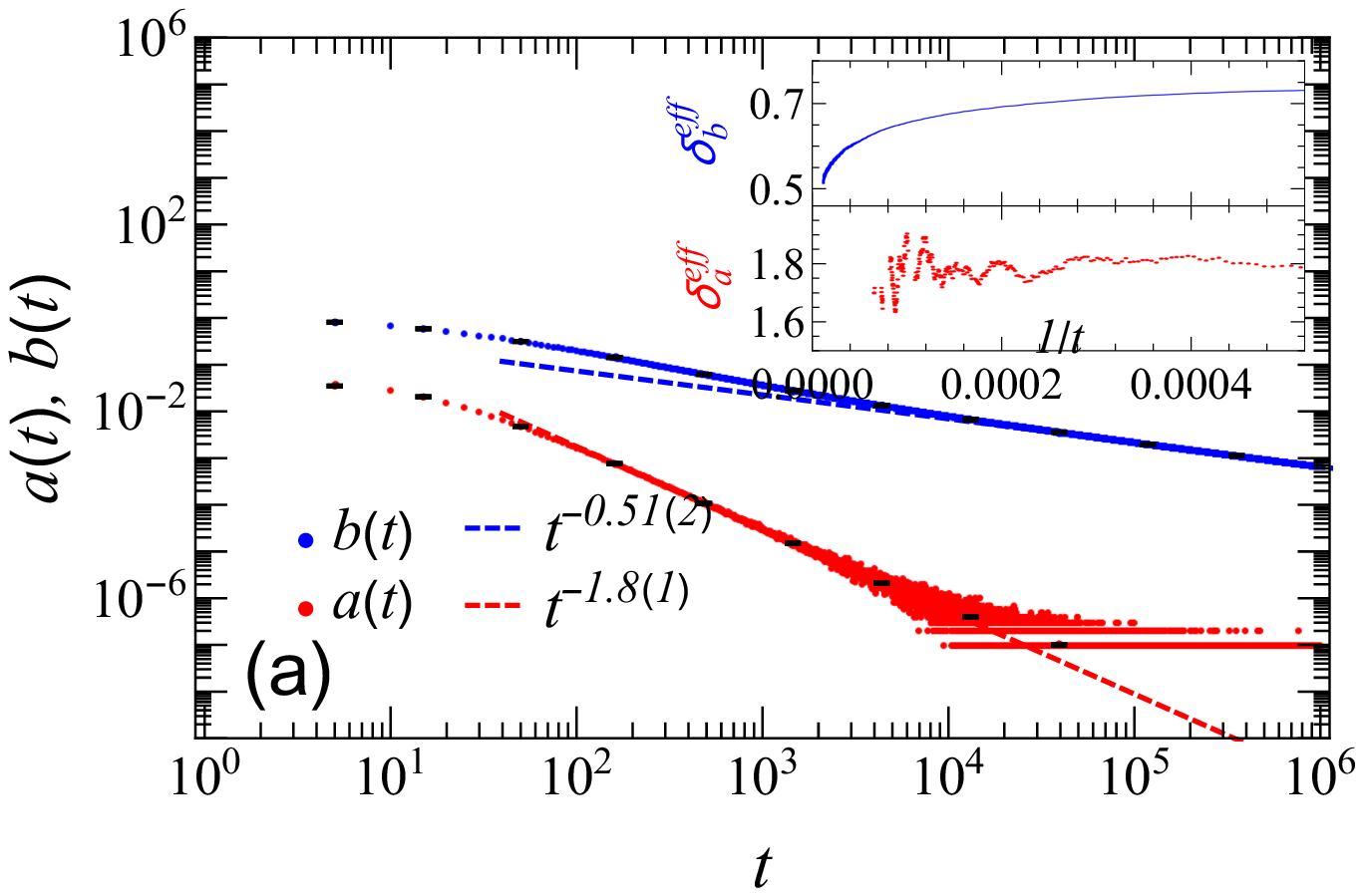}} &
  \tabbox{\includegraphics[width=0.4\textwidth]{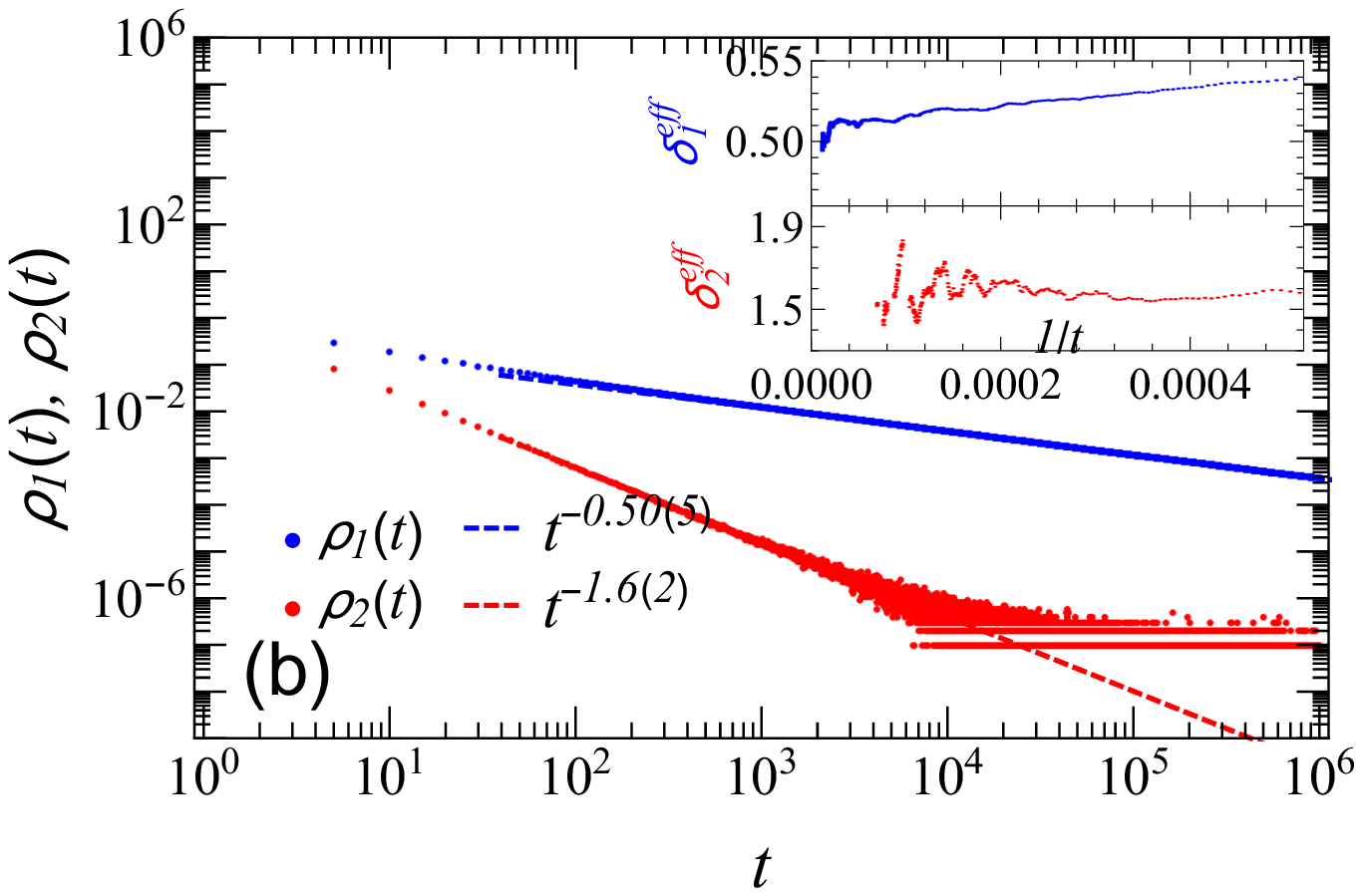}} \\
  \tabbox[c]{$d=2$} & \tabbox{\includegraphics[width=0.4\textwidth]{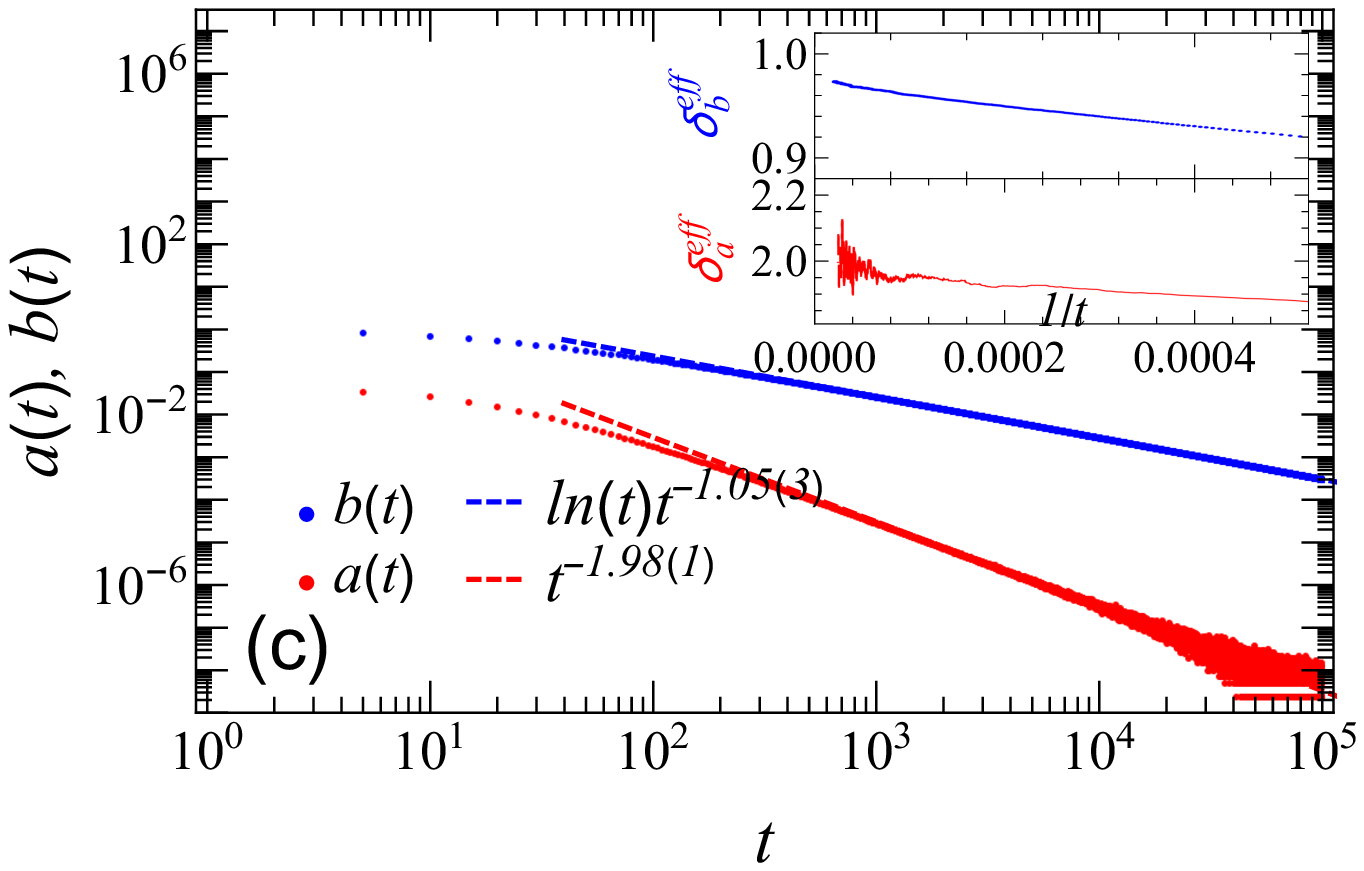}} &
  \tabbox{\includegraphics[width=0.4\textwidth]{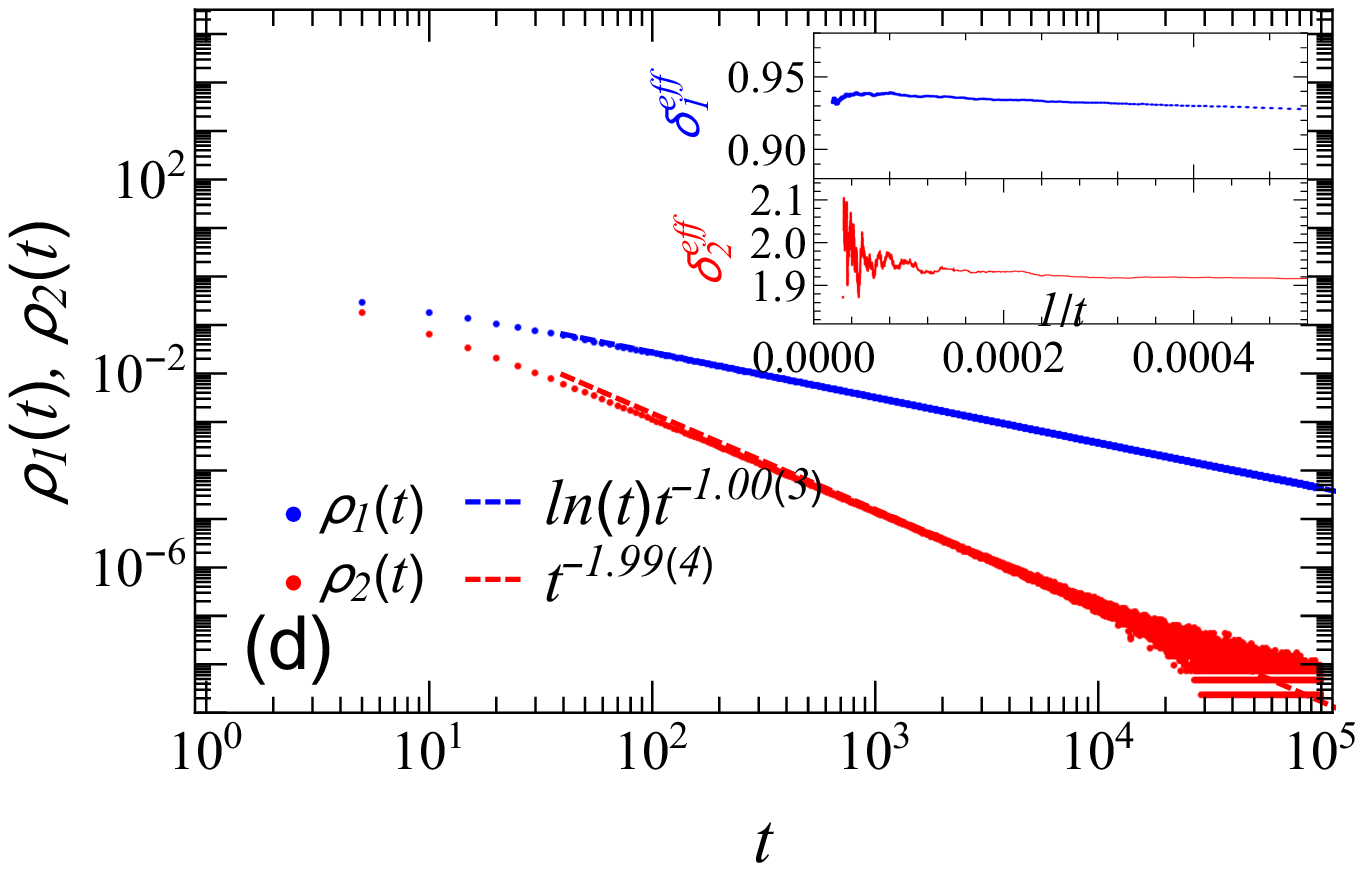}} \\
  \tabbox[c]{$d=3$} & \tabbox{\includegraphics[width=0.4\textwidth]{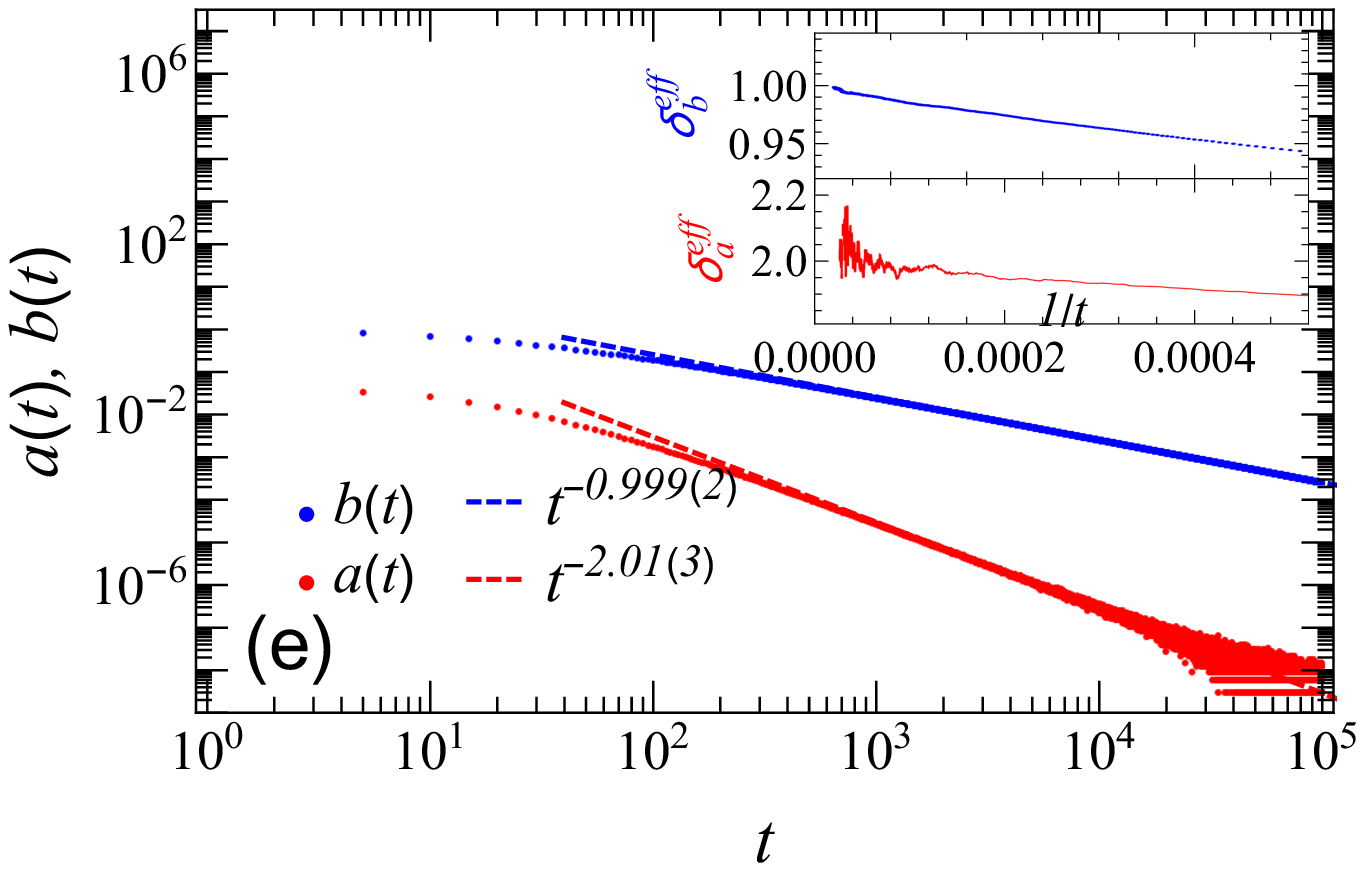}} &
  \tabbox{\includegraphics[width=0.4\textwidth]{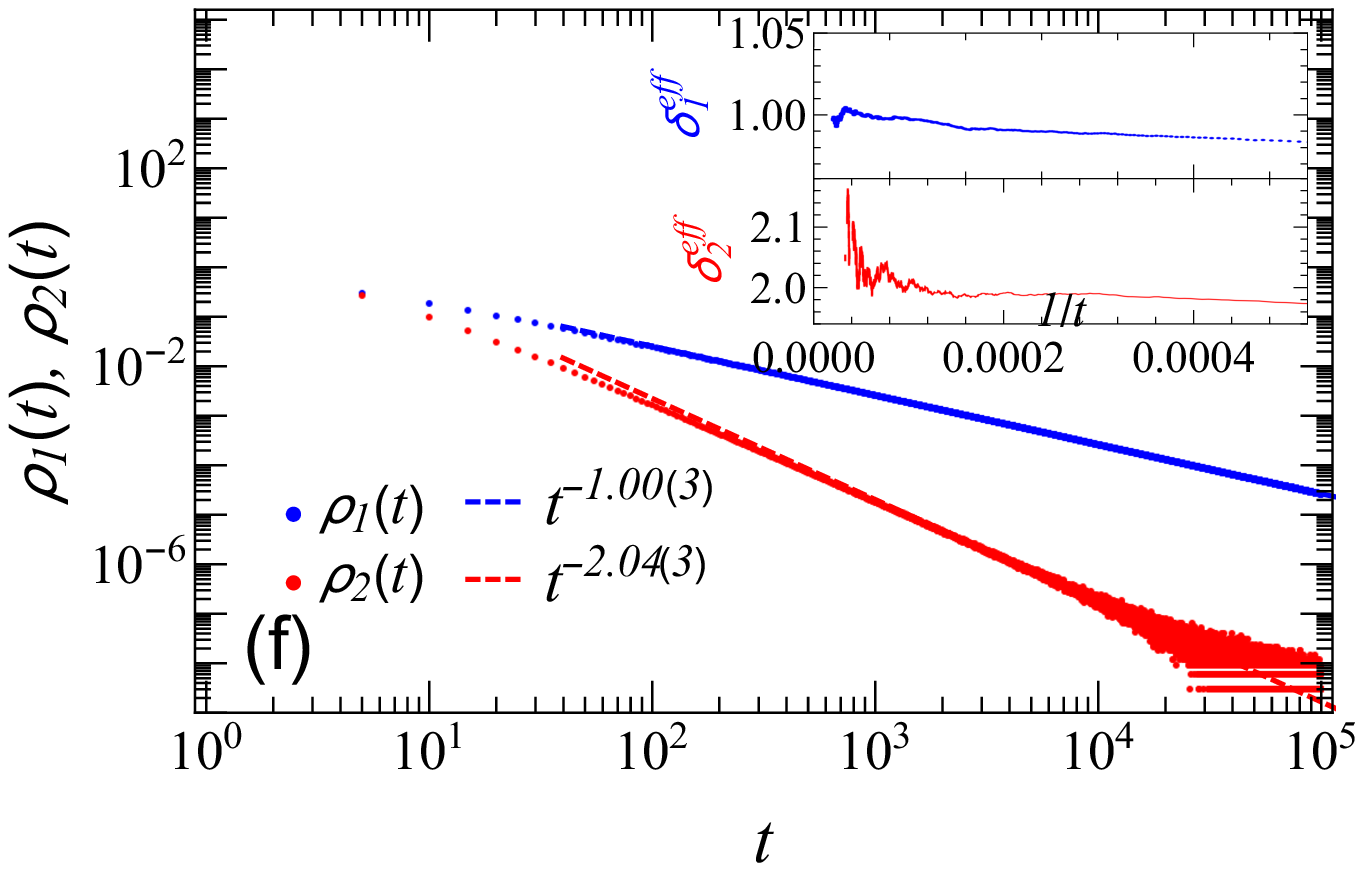}} 
  \end{tabular}
   \end{center}
   \caption{Density decay results for the inactive CPCPD [(a), (c), and (e)] and the
   inactive PCPD [(b), (d), and (f)] in one, two and three dimensions. 
   The exponents for the pair densities $a(t)$ and $\rho_2(t)$ (red), as well as the   
   exponents for the single-particle densities $b(t)$ and $\rho_1(t)$ (blue), agree well,
   within error margins, in each considered dimension. 
   The insets in each subfigure show the corresponding effective exponents. 
   All simulations started from lattices fully occupied with $B$ particles. 
   In all dimensions, the rates for the CPCPD were fixed to $D_A=0.2$, $D_B=0.22$, 
   $\mu=0.8$, $\sigma=0.6$, and $\tau=0.5$, and the rates for the PCPD were set to
   $D=0.7$ and $\mu=0.8$. 
   The other parameters were configured as follows: 
   (a) 1d CPCPD with $L=10,000$, averaged over $1,000$ independent simulation
   runs; the statistical error bars (black) for the data points are smaller than the symbol 
   size and hence omitted from the other graphs; 
   (b) 1d PCPD with $L=10,000$, averaged over $1,000$ runs; 
   (c) 2d CPCPD with $L=640$, averaged over $1,000$ runs; 
   (d) 2d PCPD with $L=640$, averaged over $1,000$ runs; 
   (e) 3d CPCPD with $L=320$, averaged over merely $10$ runs; 
   (f) 3d PCPD with $L=320$, averaged over $10$ runs.}
   \label{fig:inact}
  \end{figure*}

We initiated the simulations with our lattices fully occupied with $B$ particles. 
For the CPCPD, we measured the mean particle densities (serving as our order
parameters) $a(t)$ and $b(t)$ as defined in Eq.~\eqref{eqs:dens}. 
Denoting the number of $B$ particles at lattice site $i$ as $b_i(t) = 0,1$, there are also
two possible order parameters for the PCPD \cite{hinrichsen2001pair}, namely the 
mean single-particle density and the mean density of \textit{consecutive pairs}:
\begin{IEEEeqnarray}{rCl}
	\rho_1(t)&=&\left\langle \frac{1}{L^d}\sum_{i}b_{i}(t)\right\rangle, \nonumber \\ 
      \rho_2(t)&=&\left\langle 
	\frac{1}{2 L^d}\sum_{i}\sum_{j\in \mathrm{NN}_i} b_{i}(t) b_{j}(t) 
	\right\rangle,
	\label{eqs:rho}
\end{IEEEeqnarray}
where $\mathrm{NN}_i$ represents the set of nearest neighbors of site $i$. 
For our one- and two-dimensional systems, we averaged the quantities \eqref{eqs:dens} and 
\eqref{eqs:rho} over $1,000$ independent runs.
Our three-dimensional system sizes were of the order $320^{3} \sim 3.3 \times 10^7$,
so that averaging these observables over just $10$ runs was sufficient.

The order parameters $\rho_1(t)$ for the PCPD and $b(t)$ for the CPCPD both 
represent the single-particle ($B$) density, and should display the same scaling law. 
The second PCPD order parameter $\rho_2(t)$, albeit defined differently, should be comparable 
to the pair particle density $a(t)$ for the CPCPD. 
In principle, one may trace the consecutive pairs in the PCPD that actually annihilate or
reproduce, and regard them as `$A$ particles'; but since only the reactive pairs will be
responsible for the scaling law of $\rho_2(t)$, while the consecutive pairs that merely
split and diffuse away represent a constant background, $\rho_2(t)$ as a whole should
serve as an adequate proxy for the CPCPD order parameter $a(t)$. 
In the inactive phase, the PCPD order parameters decay algebraically as
\begin{equation}
	\rho_1(t)\sim t^{-\delta_{\mathrm{in}1}} , \quad
	\rho_2(t)\sim t^{-\delta_{\mathrm{in}2}} ,
	\label{eqs:pcpdinact}
\end{equation}
while at criticality, 
\begin{equation}
	\rho_1(t)\sim t^{-\delta_{1}} , \quad
	\rho_2(t)\sim t^{-\delta_{2}} .
	\label{eqs:pcpdcrit}
\end{equation}
If the CPCPD is equivalent to the PCPD, 
$\delta_{\mathrm{in}a} = \delta_{\mathrm{in}2}$,
$\delta_{\mathrm{in}b} = \delta_{\mathrm{in}1} = \delta_{\mathrm{pa}}$,
$\delta_a = \delta_2$, and $\delta_b = \delta_1$, where $\delta_{\mathrm{pa}}$ 
denotes the decay exponent for pure pair annihilation processes as listed in 
Eq.~\eqref{eqs2}.

\begin{figure*}[t]
   \begin{center}
\begin{tabular}{p{0.08\textwidth}p{0.44\textwidth}p{0.44\textwidth}} & 
   \multicolumn{1}{c}{CPCPD} & \multicolumn{1}{c}{PCPD} \\
   \tabbox[c]{$d=1$} & \tabbox{\includegraphics[width=0.4\textwidth]{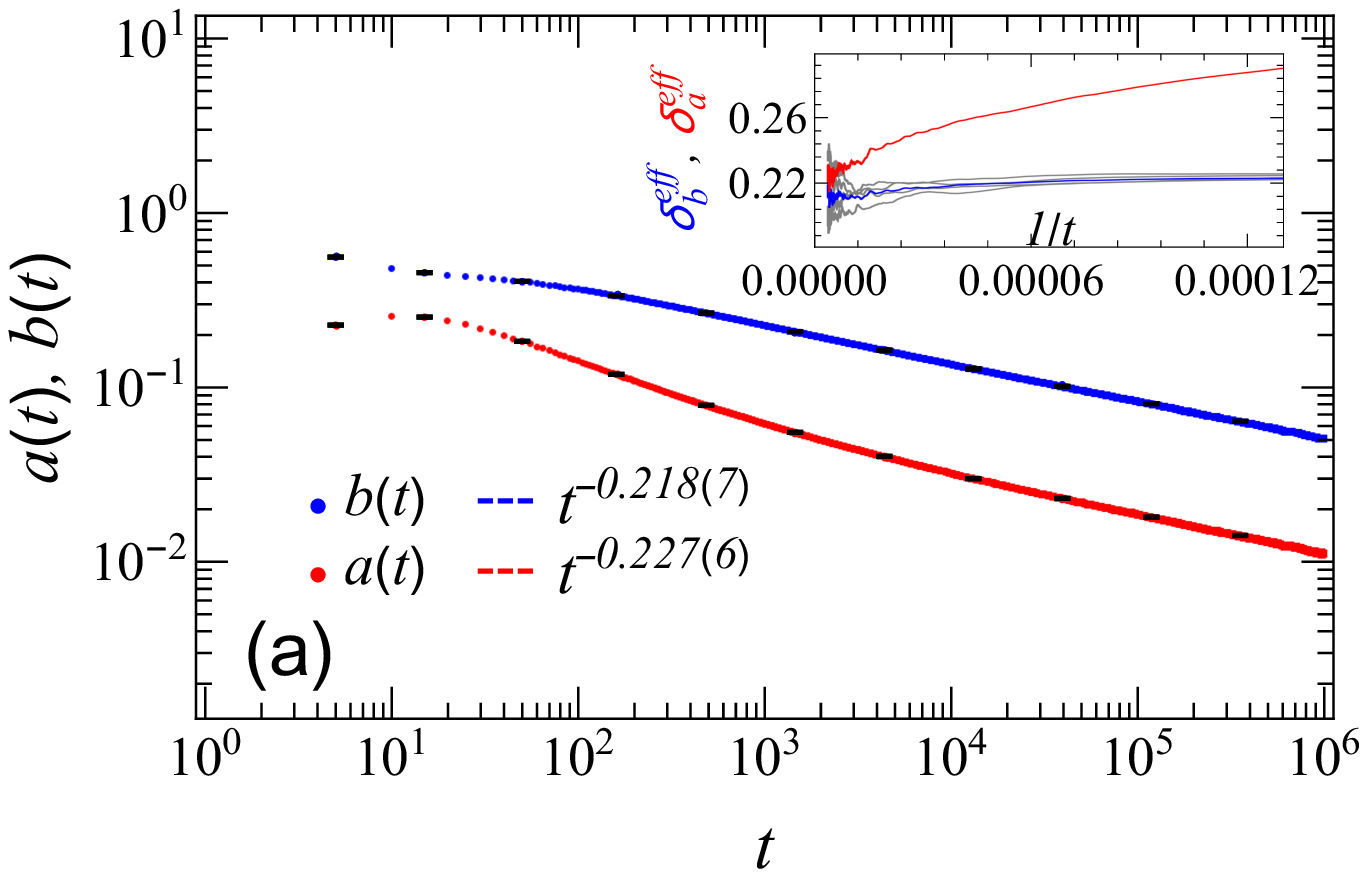}} &
   \tabbox{\includegraphics[width=0.4\textwidth]{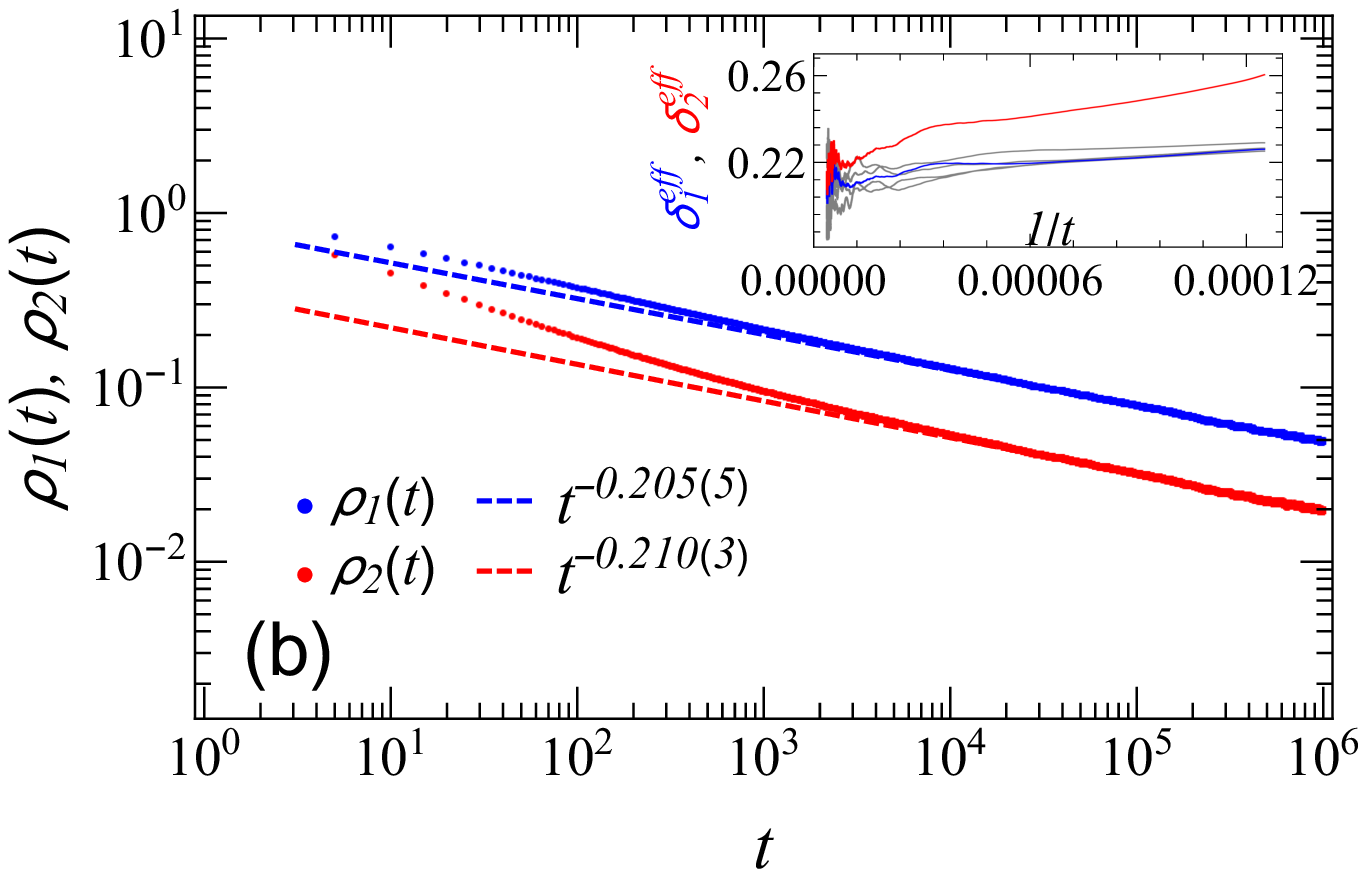}} \\
   \tabbox[c]{$d=2$} & \tabbox{\includegraphics[width=0.4\textwidth]{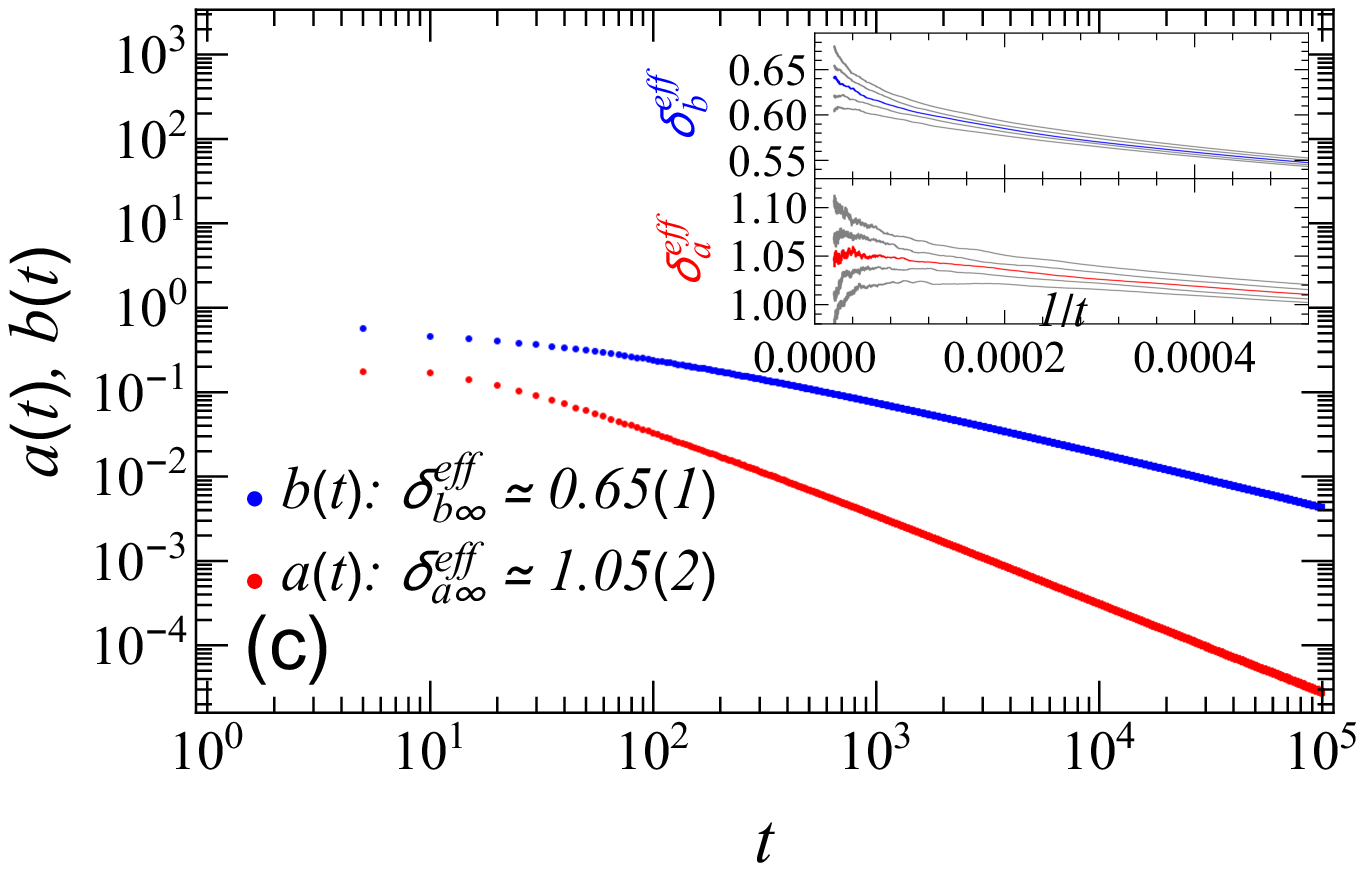}} &
   \tabbox{\includegraphics[width=0.4\textwidth]{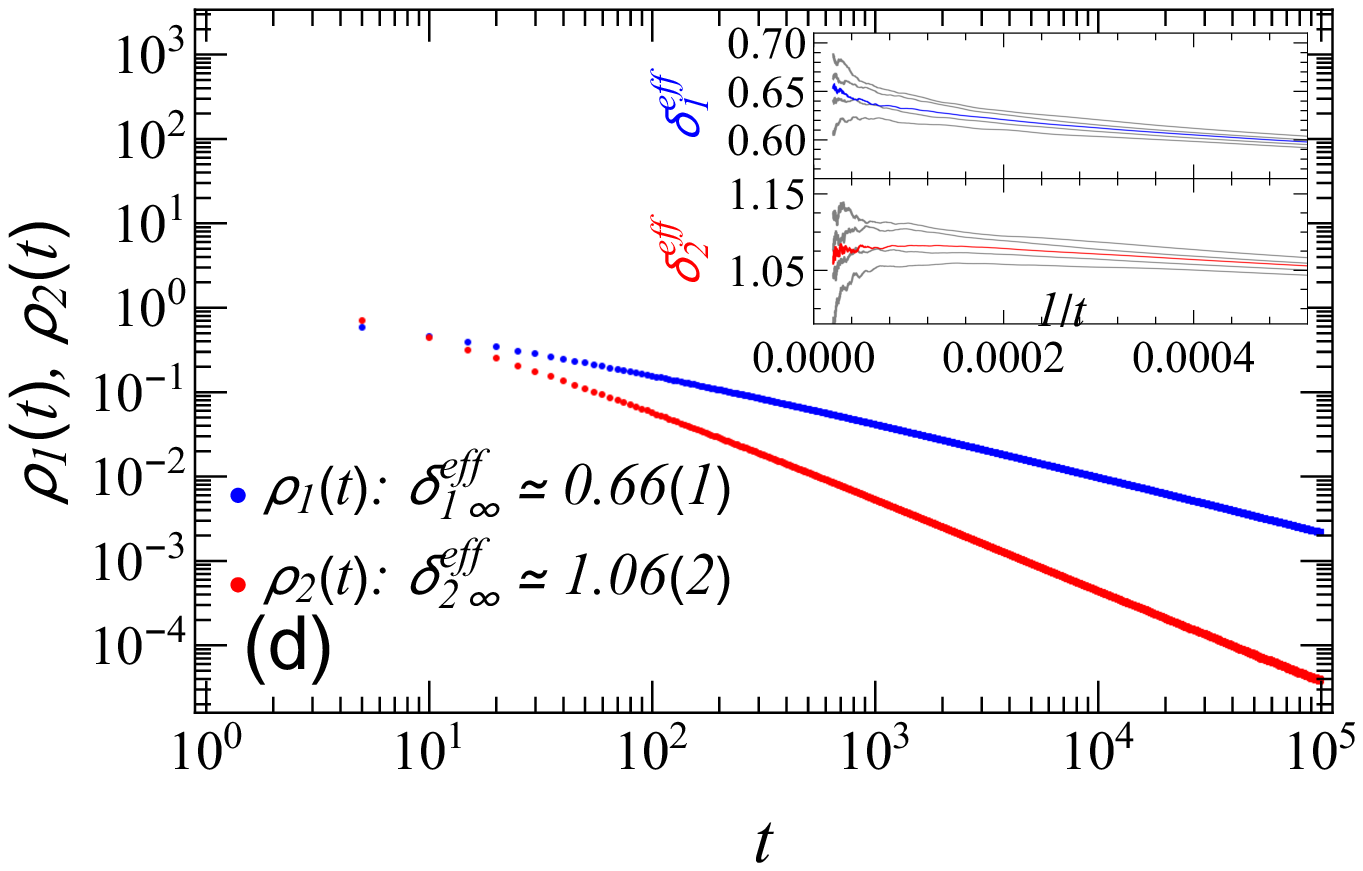}} \\
   \tabbox[c]{$d=3$} & \tabbox{\includegraphics[width=0.4\textwidth]{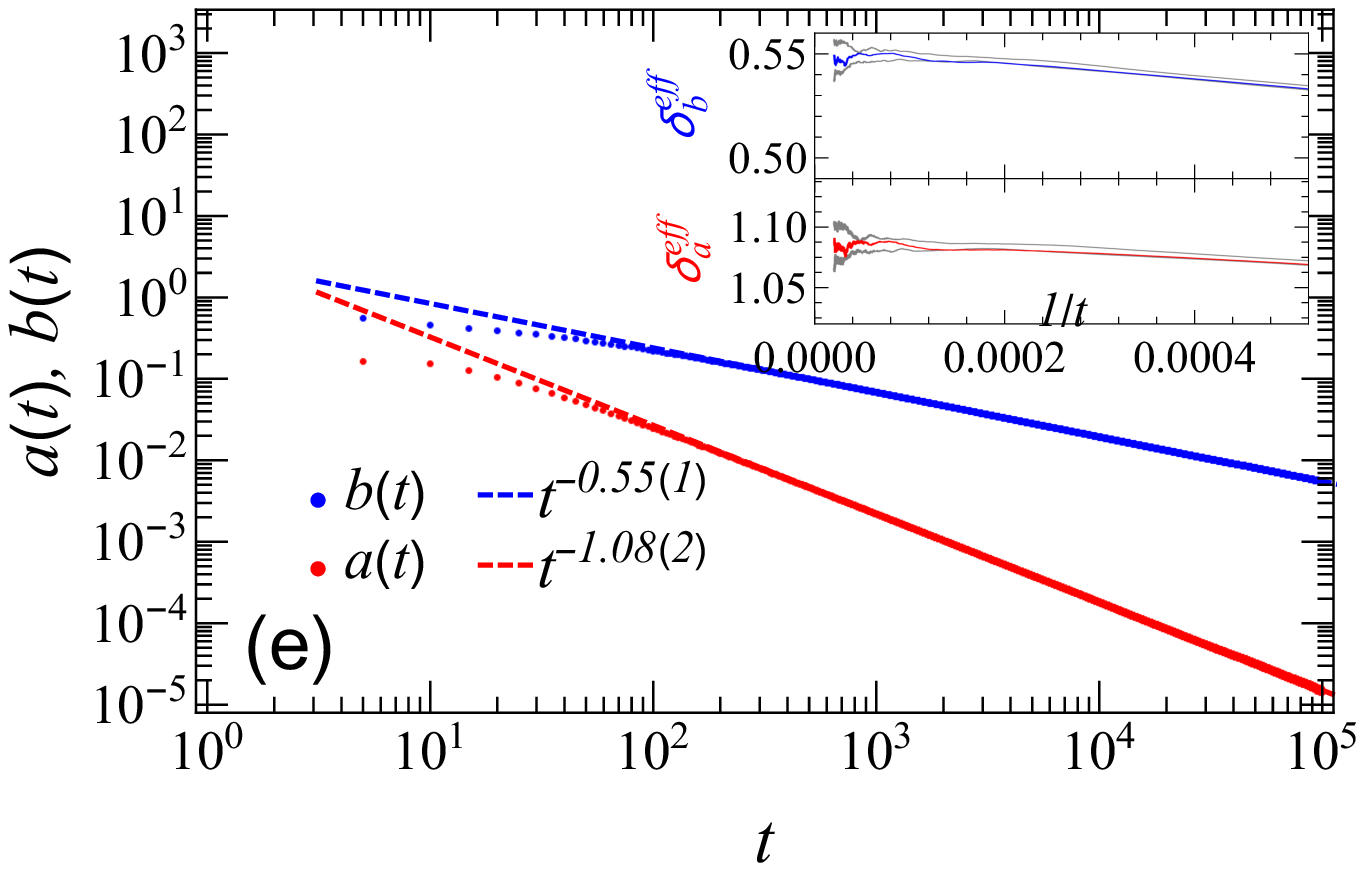}} &
   \tabbox{\includegraphics[width=0.4\textwidth]{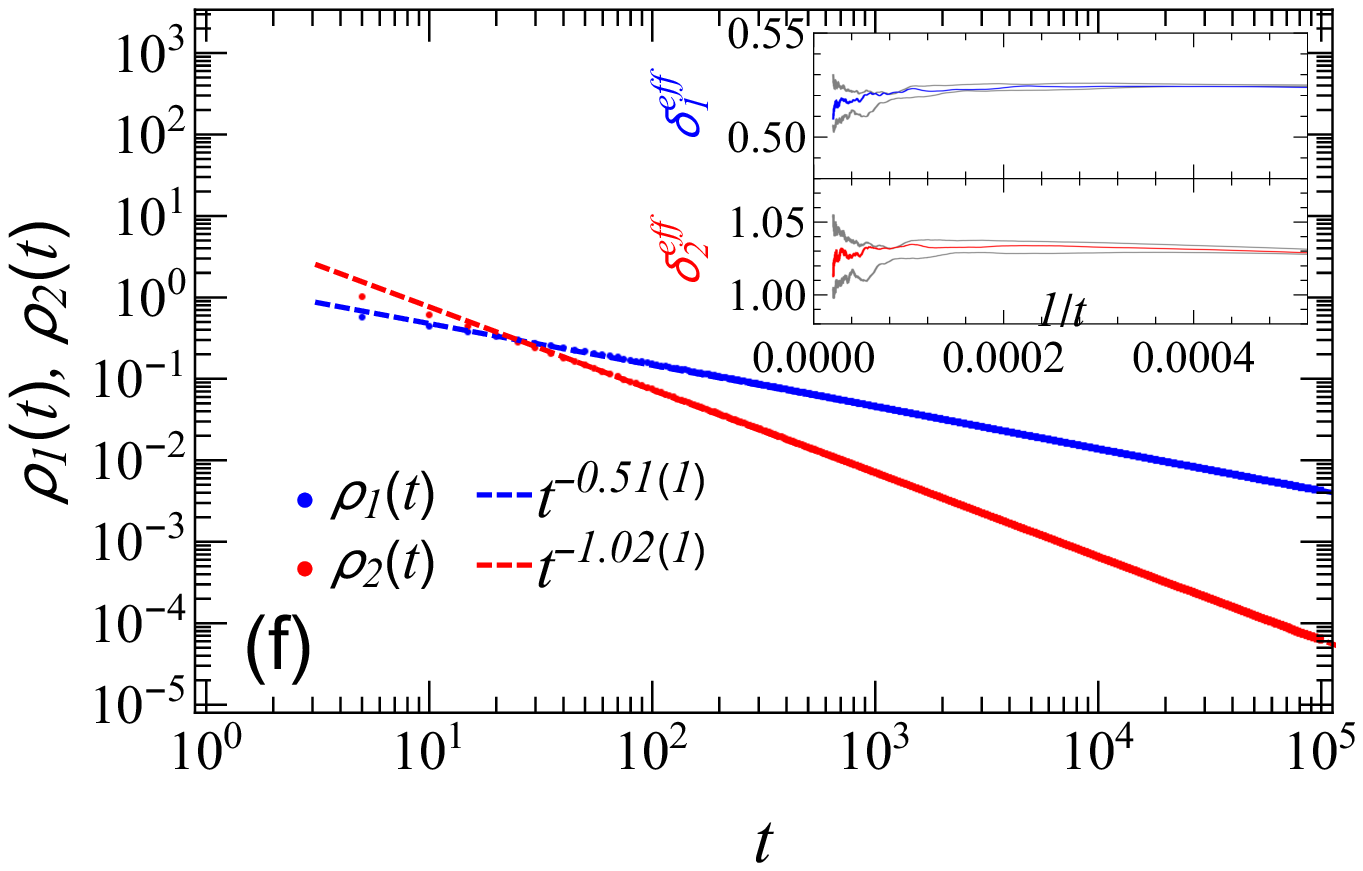}} 
   \end{tabular}
   \end{center}
   \caption{Density decay results for the critical CPCPD [(a), (c), and (e)] and the critical 
    PCPD [(b), (d), and (f)] in one, two, and three dimensions.  
   The insets in each subfigure show the corresponding effective exponents
   \eqref{eqs:eff}. 
   All simulations started from lattices fully occupied with $B$ particles. 
   In all dimensions, the rates for the CPCPD were fixed to $D_A=0.2$, $D_B=0.22$, 
   $\sigma=0.6$ and $\tau=0.5$, and the diffusion rate for the PCPD was set to 
   $D=0.7$. The other parameters and estimated critical points are as follows:
   (a) 1d CPCPD with $L=10,000$ and $\mu=0.08882$, averaged over $1,000$
   independent simulation runs; the statistical error bars (black) for the data points are 
   smaller than the symbol sizes and hence omitted for the other graphs;
   in the inset, from bottom to top, the curves pertain to
   $\mu=0.08878, 0.08880, 0.08882, 0.08884$, and $0.08886$;
   (b) 1d PCPD with $L=10,000$ and $\mu=0.15746$ (the same as in Ref.~\cite{odor2003criti}), 
   averaged over $1,000$ runs; in the inset, from bottom to top,
   $\mu=0.15742, 0.15744, 0.15746, 0.15748$, and $0.15750$; (c) 2d CPCPD with
   $L=640$ and $\mu=0.1899$, averaged over $1,000$ 
   runs; owing to superposition of the intrinsic statistical errors, corrections to scaling, 
   and the unknown logarithmic corrections at the upper critical dimension $d_c=2$, only 
   asymptotic effective exponents are given; in the inset, from bottom to top,
   $\mu=0.1895, 0.1897, 0.1899,  0.1901$, and $0.1903$; 
   (d) 2d PCPD with $L=640$ and $\mu=0.3078$, averaged over $1,000$ runs; similar
   to (c) in the inset, from bottom to top, $\mu=0.3074, 0.3076, 0.3078, 0.3080$, and    
  $0.3082$; 
  (e) 3d CPCPD with $L=320$ and $\mu=0.2134$, averaged over just $10$
  runs; in the inset, $\mu=0.2133, 0.2134$, and $0.2135$;
  (f) 3d PCPD with $L=320$ and $\mu=0.3240$, averaged over $10$ runs; 
  in the inset, $\mu=0.3239, 0.3240$, and $0.3241$}.
   \label{fig:crit}
  \end{figure*}

Setting $\mu=0.8$ placed all our systems in the inactive phase, for all dimensions 
$d = 1, 2$, and $3$.
In Fig.~\ref{fig:inact}, we plot the various densities versus time on a double-logarithmic 
scale, which allows us to infer the decay exponents via linear fits of the data.
The statistical errors for the scaling exponents were estimated via observing their 
variations as different time intervals were employed for the fits.
The CPCPD density $b(t)$ and the PCPD density $\rho_1(t)$ indeed decay according to 
the pure pair annihilation laws \eqref{eqs2}, with logarithmic corrections noticeable at 
$d_c=2$.
In agreement with the mean-field prediction, both the CPCPD $A$ particles and their 
consecutive-pair counterpart in the PCPD die out quickly with a much larger decay 
exponent in all studied dimensions: 
For $d=1$, $\delta_{\mathrm{in}a} \simeq 1.8(1) \approx \delta_{\mathrm{in}2}
\simeq 1.6(2) > \delta_{\mathrm{in}b} \simeq \delta_{\mathrm{in}1} \simeq
\delta_{\mathrm{pa}} = 1/2$; meanwhile in two and three dimensions, we observe 
the mean-field exponents $\delta_{\mathrm{in}a} \simeq 2 \simeq 
\delta_{\mathrm{in}2}$ and $\delta_{\mathrm{in}b} \simeq \delta_{\mathrm{in}1} 
\simeq \delta_{\mathrm{pa}} = 1$, c.f.~Fig.~\ref{fig:decay}. 
\begin{table}[!tbp]
       \caption{Density decay exponents for the inactive CPCPD and PCPD in one, two, 
	and three dimensions. 
	For comparison, the corresponding mean-field decay exponents (if applicable) and 
	the decay exponents for the pure pair annihilation reactions are also listed 
	[$*$ indicates logarithmic corrections as given in Eq.~\eqref{eqs2}].}
        \centering
	\begin{tabular}{r|ll|ll|c}
		\hline\hline
		& \multicolumn{2}{c|}{CPCPD} &
		\multicolumn{2}{c|}{PCPD} & pair annihil. \\
                \cline{2-6}
		& \multicolumn{1}{c}{$\delta_{\mathrm{in}a}$} 
		& \multicolumn{1}{c|}{$\delta_{\mathrm{in}b}$} 
		& \multicolumn{1}{c}{$\delta_{\mathrm{in}2}$} 
		& \multicolumn{1}{c|}{$\delta_{\mathrm{in}1}$}
		& $\delta_{\mathrm{pa}}$ \\
                \hline
                $d=1$ & 1.8(1) & 0.51(2) & 1.6(2) & 0.50(5) & 1/2 \\
                $d = d_c=2$ & 1.98(1) & $1.05(3)^{*}$ & 1.99(4) & $1.00(3)^{*}$ &
		   $1^{*}$ \\
                $d=3$ & 2.01(3) & 0.999(2) & 2.04(3) & 1.00(3) & 1 \\
                \hline
		mean-field & 2 & 1 & \multicolumn{1}{c}{--} & 1 \cite{janssen2004pair}
		& 1 \\
        \hline\hline
        \end{tabular}
 \label{tab:inactexp}
\end{table}

The algebraically slow decay of $a(t)$ and $\rho_2(t)$ indicates that at long times,
when the $B$ particle density turns low, the diffusion-limited binary reactions and the 
generation of $A$ pair particles in the CPCPD and consecutive pairs in the PCPD become 
rare events.
Consequently the $B$ particles essentially decouple from the $A$ species or consecutive
pairs, and both the CPCPD and PCPD follow pure pair annihilation kinetics in their 
inactive phases, c.f.~Table~\ref{tab:inactexp}. 
Indeed, the pure annihilation process $B+B \to \emptyset$ may be effectively realized in 
terms of the combined reactions $B+B \to A$ and $A \to \emptyset$, where an 
intermediate particle $A$ is formed whenever two $B$ particles are brought to close
proximity (in a continuum setting, within a finite reaction radius 
\cite{tauber2014critical}). 
One may then directly deduce the mean-field decay exponent for the $A$ species to be
$\delta_{\mathrm{in}a} = 2$. 
To study more systematically how the decay laws change over the system's time
evolution, we computed the effective exponents, i.e., the local slopes of the 
double-logarithmic density decay graphs $\rho(t)$,
\begin{equation}
	\delta^{\mathrm{eff}}(t) = \frac{- \ln \left[ \rho(t) / \rho(t/m) \right]}{\ln (m)} 
      	\, ;
	\label{eqs:eff}
\end{equation}
we used $m=8$.
The insets of Fig.~\ref{fig:inact} show that the effective density decay exponents 
approach their asymptotic values slowly; this crossover is delayed for the CPCPD owing 
to the intermediate pair production processes. 

Particularly in three dimensions, the CPCPD and PCPD crossover features are in 
accord with the mean-field prediction, see Fig.~\ref{fig:mfeffexp}(a) in 
App.~\ref{appda}.
Note that in one dimension, according to the correlation of next-neighbor particles 
\cite{alcaraz1994re}, the pair density for the pure pair annihilation process should scale
as $\rho_2(t)\sim t^{-3/2}$, which also holds for the inactive one-dimensional PCPD 
\cite{park2005driven}. 
Yet the measured value $\delta_{\mathrm{in}a} \simeq 1.8(1)$ for the one-dimensional
CPCPD is somewhat larger.
This discrepancy too might be ascribed to the slower crossover behavior of the CPCPD. 
Futhermore, the rather noisy tail for $a(t)$ causes larger measurement errors.

\begin{table*}[!tbp]
       \caption{Critical point locations and critical density decay exponents for the CPCPD 
	and PCPD in one, two, and three dimensions. 
	For comparison, the corresponding mean-field critical 	exponents (if applicable) and
 	the reported critical exponents for the PCPD are also displayed [$**$ indicates the
	asymptotic effective exponent values, see Eq.~\eqref{eqs:eff}, at large $t$]. 
      According to recent studies \cite{hinrichsen2006phase, park2014critical}, a 
      more precise range for $\delta_1$ in one dimension should be $0.17$--$0.185$.}
        \centering
        \begin{tabular}{r|lll|lll|l}
		\hline\hline
		& \multicolumn{3}{c|}{CPCPD} &
		\multicolumn{3}{c|}{PCPD (this work)} &
		\multicolumn{1}{c}{PCPD} \\
                \cline{2-8}
		& \multicolumn{1}{c}{$\mu_c$} &
		\multicolumn{1}{c}{$\delta_{a}$} &
		\multicolumn{1}{c|}{$\delta_{b}$} &
		\multicolumn{1}{c}{$\mu_c$} & 
		\multicolumn{1}{c}{$\delta_{2}$} & 
		\multicolumn{1}{c|}{$\delta_{1}$} &
		\multicolumn{1}{c}{$\delta_{1}$}\\
                \hline
		$d=1$ & 0.08882(1) &  0.218(7) & 0.227(6) &
		0.15746(1)
	        & 0.210(3) &
		0.205(5) & 0.17--0.25
		\cite{henkel2004non,smallenburg2008univ} \\
		$d = d_c = 2$ & 0.1899(1) & 1.05(2)$^{**}$ &
		0.65(1)$^{**}$  &
		0.3078(1) &
		1.06(2)$^{**}$ & 0.66(1)$^{**}$  & 0.5 ($\simeq\delta_2$
   		\cite{odor2002phase}) \\
             $d=3$ & 0.2134(1) & 1.08(2) & 0.55(1) & 0.3240(1) & 1.02(1) & 
		0.51(1) & \multicolumn{1}{c}{--} \\
                \hline
		mean-field  & \multicolumn{1}{c}{--} & 1 & 1/2 &
		\multicolumn{1}{c}{--} & \multicolumn{1}{c}{--} & 1/2
		& \multicolumn{1}{c}{1/2}\\
        \hline\hline
        \end{tabular}
        \label{tab:critexp1}
\end{table*}
Examining either system at criticality requires precise estimates of $\mu_c$, which we
achieved by analyzing the large-time local slopes in the density decay curves. 
The details of our simulation setups are described in the caption of Fig.~\ref{fig:crit}. 
In one dimension, c.f.~Figs.~\ref{fig:crit}(a) and (b), similar to the earlier observation 
for the PCPD that $\delta_1\simeq \delta_2$ \cite{hinrichsen2001pair}, the CPCPD gives 
$\delta_a\simeq \delta_b\approx \delta_1\simeq \delta_2$. 
The apparently larger critical decay exponents for the CPCPD as compared to the PCPD
may be ascribed to its slower crossover dynamics, since the effective exponents 
$\delta^{\mathrm{eff}}(t)$ of the latter show a slow downward drift with increasing 
time \cite{hinrichsen2006phase}, and the same tendency should also be expected for 
the CPCPD. 
Therefore at least in one dimension, the numerically determined decay exponents for 
the CPCPD turn out slightly larger than those of the PCPD obtained at comparable time 
scales. 
Our measured exponents for the one-dimensional CPCPD also reside in the range 
reported in the literature, as listed in Table~\ref{tab:critexp1}. 

Field-theoretic analysis \cite{howard1997real,janssen2004pair} and early simulations
\cite{odor2002phase} predicted that the upper critical dimension of the PCPD is 
$d_c  = 2$. 
Indeed, in two dimensions, c.f.~Figs.~\ref{fig:crit} (c) and (d), we observe that the 
effective exponents of $a(t)$ and $\rho_2(t)$ become stationary at large $t$ and
approach the mean-field CPCPD value $\delta_a=1$, whereas the corresponding 
effective exponents for $b(t)$ and $\rho_1(t)$ keep veering up until they both approach
the value $\delta_b^{\mathrm{eff}} \simeq \delta_{1}^{\mathrm{eff}} \simeq 0.65$, 
indicating that there exist nontrivial corrections to scaling which induce deviations of 
the exponent values away from their mean-field expectation $\delta_b = \delta_1 = 1/2$. 
This is of course to be anticipated at the critical dimension $d_c$, where for the 
(C)PCPD as yet unknown logarithmic corrections may superpose the intrinsic corrections 
to scaling. 
It appears that the critical pair density $a(t)$ for the CPCPD and likewise $\rho_2(t)$ for
the PCPD are plagued by such scaling corrections to a lesser extent than the 
single-particle densities $b(t)$ and $\rho_1(t)$, at least when particle number 
parity conservation is not imposed. 

We note that in stark contrast to our results, in their study of the two-dimensional 
parity-conserving PCPD, \'{O}dor, Marquez, and Santos obtained 
$\delta_1 = \delta_2 \simeq 0.5$ after employing correction terms in the form of certain
combinations of powers of logarithms \cite{odor2002phase}. 
A similar conclusion was also drawn for the one-dimensional driven PCPD 
\cite{park2005driven}, where the critical dimension was considered to be reduced by one.
To clarify this discrepancy, we have devoted a separate study for the critical decay 
exponents of the parity-conserving PCPD, summarized in App.~\ref{appdb}. 
It appears that particle number parity conservation indeed plays a significant role in two 
dimensions, as it modifies the strong correction-to-scaling contributions.

Beyond these subtle discrepancies in two dimensions that may be caused by large 
corrections to scaling, both our CPCPD and PCPD data for three-dimensional systems,
shown in Figs.~\ref{fig:crit} (e), (f) and Fig.~\ref{fig:3dpar}, clearly display the 
mean-field CPCPD critical decay exponents $\delta_a \simeq \delta_2 \simeq 1$ and 
$\delta_b \approx \delta_1 \approx 1/2$, confirming that $d_c < 3$ for both models. 
The crossover features  of the effective exponents at large times apparently 
resemble those of Fig.~\ref{fig:mfeffexp} (b) for the mean-field CPCPD.

To summarize and close this subsection, we have collected the estimated density decay
exponents in the inactive phase in Table~\ref{tab:inactexp}, and their critical 
counterparts in Table~\ref{tab:critexp1}. 
Within statistical and systematic error margins, the corresponding order parameters for 
the CPCPD and the PCPD display the same scaling properties, both in the absorbing 
state phase and at criticality.
In the inactive phase, both the CPCPD and the PCPD are described by pure 
diffusion-limited pair annihilation processes. 
The mean-field exponents for the CPCPD are recovered for $d \ge 2$ for both models,
except for $\delta_b$ and $\delta_1$ at the critical dimension $d_c = 2$. 
Attributing these deviations to (logarithmic) corrections to scaling, our data yield critical
decay exponents in two dimensions that appear consistent with those of 
three-dimensional systems and the mean-field CPCPD, strongly suggesting that the 
upper critical dimension of the critical (C)PCPD is indeed $d_c=2$.

\subsection{Seed simulations \label{sec3.3}}

Clusters generated from a single seed provide another important means to explore the
dynamical critical properties of continuous phase transitions \cite{grassberger1979re}. 
In standard active to absorbing phase transitions where single-particles are able to 
reproduce, the growth of clusters is characterized by the survival probability
$P_{\mathrm{sur}}(t)$ up to time $t$; the number of active sites $N(t)$ at $t$, with 
both quantities averaged over all runs; and the mean square spreading from the origin,
$R(t)^2 = \left\langle \sum_{i} s_i(t) |\mathbf{r}_i|^2 / N(t) \right\rangle$, averaged
over surviving clusters, where $s_i(t)=0,1$, and $\mathbf{r}_i$ denotes the 
displacement from the origin to site $i$. 
At criticality, the dynamics becomes scale invariant and these quantities follow the power
laws
\begin{eqnarray}
	P_{\mathrm{sur}}(t) \sim t^{-\delta'} , \quad 
	N(t) \sim t^{\Theta} , \quad R(t)^2 \sim t^{\tilde{z}} ,
	\label{eqs:seedone}
\end{eqnarray}
with the the survival probability exponent $\delta'$, initial slip exponent $\Theta$, and
the spreading exponent $\tilde{z} = 2/z$ in standard seed simulations. 
For the PCPD, one of course needs to start from a localized pair of particles, which 
poses a conceptual problem since the pair-connectedness function is not well-defined
\cite{henkel2008non}; nevertheless the above power laws appear to hold for
single-particle statistics \cite{hinrichsen2003stoc}, provided at least two particles remain
in the system, and the seeded stochastic process hence stays active.

\begin{figure*}[!hptb]
      \begin{center}
		\includegraphics[width=0.82\textwidth]{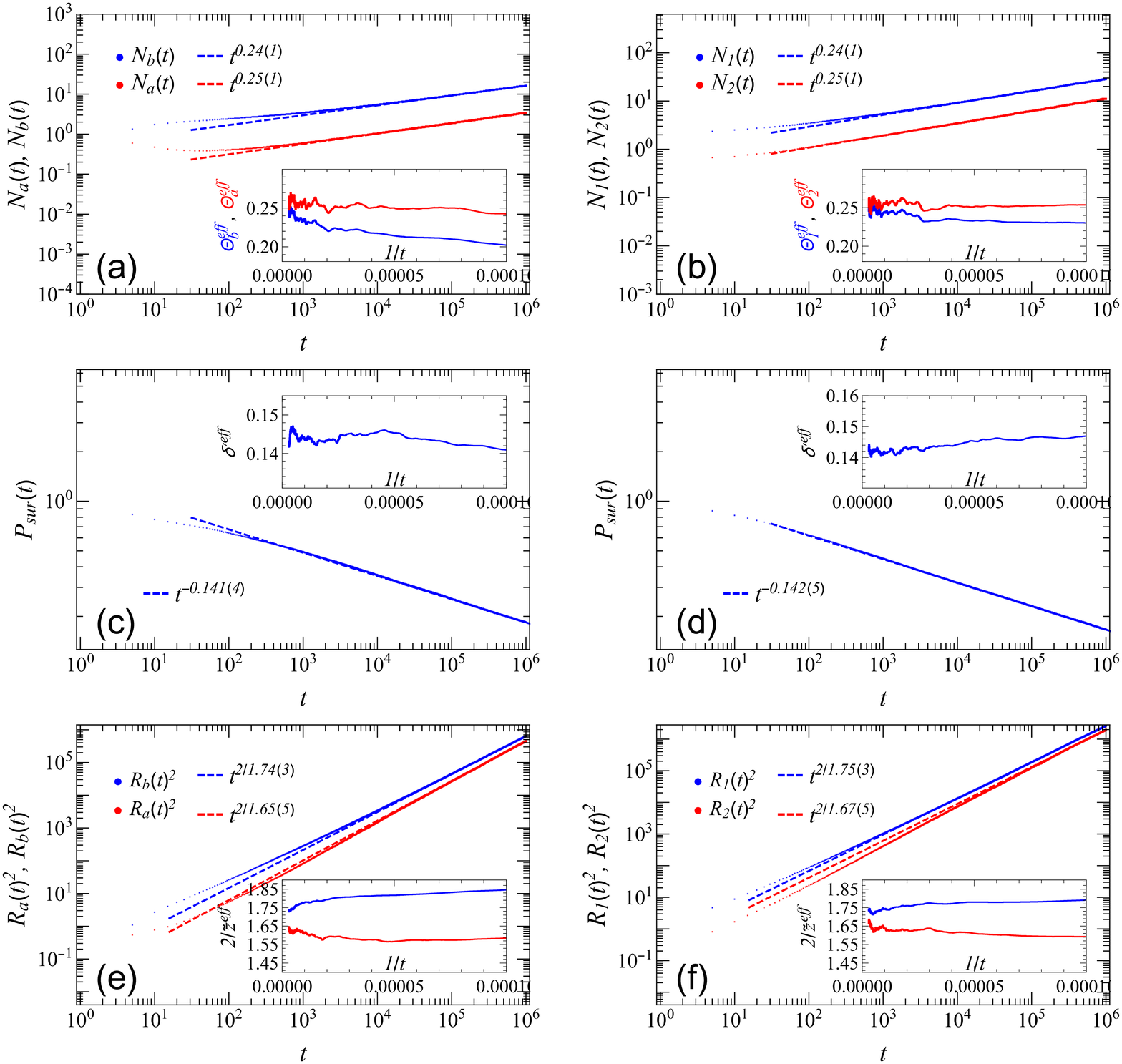}
      \end{center}
      \caption{Seed simulation results at criticality for (a), (c), (e) the one-dimensional
	      CPCPD, with $\mu_c \simeq 0.088785(5)$, $D_A = 0.2$, $D_B = 0.22$, 
	$\sigma = 0.6$, and $\tau = 0.5$; (b), (d), (f) the one-dimensional PCPD, with 
	$\mu_c\simeq 0.15744(1)$, $D=0.7$; for both systems, $L = 10,000$, and the
	data were averaged over $50,000$ independent Monte Carlo runs.}
       \label{fig:seed}
\end{figure*}

For the CPCPD, with two species at our disposal, seed simulations naturally start with a
single pair-particle $A$. 
The ensuing spreading process continues as long as at least one $A$ or two $B$
particles survive. 
The power laws \eqref{eqs:seedone} are readily extended to 
\begin{eqnarray}
	P_{\mathrm{sur}}(t) \sim t^{-\delta'}, \quad 
	& N_a(t) \sim t^{\Theta_a} , \quad & R_a(t)^2 \sim t^{\tilde{z}_a} , \quad \\
      & N_b(t) \sim t^{\Theta_b} , \quad & R_b(t)^2 \sim t^{\tilde{z}_b} , \quad
	\label{eqs:seedcpcpd}
\end{eqnarray}
where the subscripts ``$a$'' and ``$b$'' denote the corresponding quantities and
scaling exponents for species $A$ and $B$, respectively. 
Figures \ref{fig:seed} (a), (c), and (e) show the seed simulation results for the CPCPD
on a one-dimensional lattice. 
The critical point $\mu_c$ was determined by observing the effective exponents for
$N_a(t)$ and $N_b(t)$, see the inset of Fig.~\ref{fig:seed} (a). 
The critical point for seed simulations seems to shift slightly from 
$\mu_c \simeq 0.08882(1)$ for the initially fully occupied lattice to
$\mu_c \simeq 0.088785(5)$ for a single initial seed. 
Since finite-size effects are eliminated in seed simulations so that an upward shift
of $\mu_c$ should be expected if finite-size effects were the cause, this observed 
downward shift rather hints again at strong correction to scaling since couplings 
between the two species occur much less frequently than in homogeneous initial 
configurations.

We observe that the exponents $\Theta_a$ and $\Theta_b$, as well as $\tilde{z}_a$ 
and $\tilde{z}_b$, take very close numerical values. 
For direct comparison, Figs.~\ref{fig:seed} (b), (d), and (f) display the seed simulation 
results for the one-dimensional PCPD, for which the estimated critical point is also 
shifted slightly. 
Here we account for the corresponding quantities for consecutive pairs as well, denoting
them with the subscript ``2'', while the corresponding single-particle observables are 
labeled with the subscript ``1''. 
Hence for the PCPD, one expects the critical seed scaling
\begin{eqnarray}
	P_{\mathrm{sur}}(t) \sim t^{-\delta'} , \quad 
	& N_1(t) \sim t^{\Theta_1} , \quad & R_1(t)^2 \sim t^{\tilde{z}_1} , \quad \\
      & N_2(t) \sim t^{\Theta_2} , \quad & R_2(t)^2 \sim t^{\tilde{z}_2} . \quad
	\label{eqs:seedpcpd}
\end{eqnarray}

We take the close critical exponent values measured in the CPCPD and PCPD seed
simulations, namely 
$\delta'_{\mathrm{CPCPD}}\simeq 0.14 \simeq \delta'_{\mathrm{PCPD}}$,
$\Theta_a\simeq \Theta_2 \approx 0.25 \approx \Theta_b \simeq \Theta_1$,
$2/\tilde{z}_a \simeq 1.65 \simeq 2/\tilde{z}_2$, and 
$2/\tilde{z}_b \simeq 1.75 \simeq 2/\tilde{z}_1$, as evidence that the nonequilbrium
phase transitions in these distinct nonlinear stochastic reactions are governed by the
same universality class, i.e., the scaling properties of the CPCPD are fully consistent with
those of the PCPD, and demonstrate the relevance of consecutive pairs in the PCPD.

Since in standard seed simulations $2/\tilde{z}=z$ holds, it is tempting to interpret 
$2/\tilde{z}_a$ and $2/\tilde{z}_2$, respectively, as the corresponding dynamical 
exponents for species $A$ and $B$. 
For our CPCPD and PCPD simulation, we provide the effective exponents for 
$2/\tilde{z}_a$ and $2/\tilde{z}_2$ as well as $2/\tilde{z}_b$ and $2/\tilde{z}_1$ in 
the insets of Figs.~\ref{fig:seed} (e) and (f). 
Indeed, our obtained asymptotic exponent values $2/\tilde{z}_b \simeq 2/\tilde{z}_1 
\simeq 1.75$ reside in the reported range of the conventionally measured dynamical 
exponent $z$ of the PCPD in the literature.
Table~\ref{tab:critexp} list a full comparison of the critical exponents of different 
models in one dimension, demonstrating that the critical properties of the CPCPD are
fully consistent with those of the PCPD, and markedly differ from those of the DP and 
the PC universality classes.
\begin{table}[!bp]
 \centering
 \caption{Seed simulation exponents for the PCPD, CPCPD, DP, and PC universality
	classes. 
	In conventional active to absorbing phase transitions,  $z = 2 / \tilde{z}$.
 	The subscripts for the quantities defined in this paper are indicated, where 
	applicable, in the second column.}
 \begin{tabular}{lcllll}
 \hline\hline
 Critical exponents  &  & \multicolumn{1}{c}{$\delta'$} & 
 \multicolumn{1}{c}{$\Theta$} & \multicolumn{1}{c}{$2/\tilde{z}$} \\
 \hline
 PCPD \cite{henkel2004non,smallenburg2008univ} & -- &  0.09-0.15 \, & 0.1-0.23 \,
 & 0.17-2.0 \\
 \hline
 \multirow{2}{*}{PCPD \cite{kwon2007double}} & -- & 0.130 & 0.275 & $1.61(1)$ \\
 			& -- &      &       & $1.768(8)$ \\
 \hline
 \multirow{2}{*}{CPCPD} & \, a \, & 
 \multirow{2}{*}{0.141(4)} & 0.25(1) & 1.65(5) \\
              	& b  &  & 0.24(1) & 1.74(3) \\
 \hline
 \multirow{2}{*}{PCPD (this work)} \ & 2 & \multirow{2}{*}{0.142(5)} & 0.25(1) 
  			& 1.67(5) \\
                   & 1 &   & 0.24(1) & 1.75(3)\\
 \hline
 DP \cite{jensen1999low} & -- &  0.1595     & 0.3137    & 1.5807 \\
 PC \cite{park2013high}  & -- &  0   & 0.2872   &
 1.7415  \\
 \hline\hline
 \end{tabular}
 \label{tab:critexp}
\end{table}

Adopting the CPCPD as an apt representation of the PCPD universality class, we are now 
in a position to commence analysis of the seed simulation exponents $\Theta_a$, 
$\Theta_b$, $\tilde{z}_a$, and $\tilde{z}_b$. 
It is customary to introduce two pair-connectedness functions, 
$\Upsilon_{aa}(t, \mathbf{r}; \Delta)$ and $\Upsilon_{ab}(t, \mathbf{r}; \Delta)$ 
(see App.~\ref{appdc}) that represent the probabilities of finding an $A$ or $B$ 
particle at space-time coordinate $(t, \mathbf{r})$, in a sequence of stochastic
processes starting from a seed particle $A$ located at $\mathbf{r}=0$ at time $t=0$. 
With an initial $A$ seed and two distinct pair-connectedness functions, the conceptual 
difficulty residing in seed simulations for the original PCPD is naturally resolved. 
As remarked earlier in section \ref{sec3.2}, earlier studies were already able to obtain 
consistent values for certain critical exponents  for the one-dimensionsal PCPD
\cite{henkel2004non,odor2003criti}. 
However, the measured exponents that relate to correlations, such as the dynamical exponent, 
displayed noticeable dependences 
on the diffusion rate, suggesting more complex dynamical critical behavior. 
The CPCPD is obviously governed by multiple length (and time) scales,
namely 
$\xi_a$ ($t_{c a}$), $\xi_b$ ($t_{c b}$), $\xi_{ab}$ ($t_{c ab}$) and $\xi_{ba}$ 
($t_{c ba}$) indicating the (cross-)correlation lengths (and characteristic relaxation 
times) for the two species on the (infinite) seed cluster, which in turn are related to the 
cutoffs of the correlation (autocorrelation) functions $\langle a_i(t) a_{i+r}(t) \rangle$ 
[$\langle a_i(t) a_{i}(t+\Delta t) \rangle$], $\langle b_i(t) b_{i+r}(t) \rangle$ 
[$\langle b_i(t) b_{i}(t+\Delta t) \rangle$], $\langle a_i(t) b_{i+r}(t) \rangle$ 
[$\langle a_i(t) b_{i}(t+\Delta t) \rangle$], and $\langle b_i(t) a_{i+r}(t) \rangle$
[$\langle b_i(t) a_{i}(t+\Delta t) \rangle$]. 
In addition,  the solitarily diffusing particles mark another growing length scale 
$R_{b0}(t) \sim t^{1/2}$ and the characteristic time $t_{c b0}\sim L^{2}$, with 
diffusive dynamical exponent $2$ (in accord with previous field theory analyses
\cite{howard1997real,janssen2004pair}). 

We surmise that the strong corrections to the asymptotic critical scaling in the (C)PCPD
should be attributed to the competition between the various length and time scales in 
the system, until eventually all other length (time) scales are surpassed by one dominant 
length (time) scale that characterizes the infinite spreading cluster and hence its ultimate 
scale-invariant properties. 
At the relatively short time scales that are usually accessed in numerical simulations, the
various (auto)correlation functions may still feature distinct (cross-)correlation lengths 
(times), all of which play some roles in the short-time dynamics, leading to slow 
crossover behavior that moreover varies for different observables.
However, at sufficiently (here likely extremely) long time and large length scale 
(pushing $r \to \infty$ and $\Delta t \to \infty$ to probe the corresponding correlation 
lengths and characteristic times), all (auto-)correlation functions describe the 
correlations between a pair of particles connected by a sequence of numerous 
intermediate detached diffusive $B$ particles and reactive $A/B$ particle clusters.
In ensemble averages, these distinct correlations will ultimately be reigned by the 
largest correlation length and time, and yield identical scaling exponent values in the
limits $r \to \infty$ and $\Delta t \to \infty$. 
In particular, the purely diffusive scales $R_{b0}(t)$ and $t_{c b0}$ are rendered
irrelevant in the asymptotic critical regime (as opposed to the inactive, absorbing phase)
because the chance of two particles being correlated via mere diffusive $B$ particles 
becomes exceedingly rare.

In mean-field theory, the dominant length and time scales are $\xi_a$ and $t_{c a}$ 
(c.f.~Sec.~\ref{sec2.2}). 
In one dimension, we should also expect $\xi_a$ and $t_{ca}$ to govern the system's
critical scaling properties, according to the observed inequality sequence
$2 / \tilde{z}_a < 2 / \tilde{z}_b < 2$ [see App.~\ref{appdc} and Eq.~\eqref{eqs:a13} for 
a more detailed discussion of their physical meaning], indicating that the $A$ pair 
particles spread superdiffusively as $R_a(t) \sim t^{\tilde{z}_a}$ with the largest 
spreading exponent to create the longest correlation length. 
This of course reflects the fact that all CPCPD activity requires the presence of pair 
particles $A$, whence their mutual correlations should be prevalent in the system.
Indeed, this is in agreement with the observation in 
Refs.~\cite{dickman2002nonun,de2006moment} that the subspace with at least one particle 
pair is responsible for the scaling features in the PCPD.
By postulating this dominant length and time scale and with the aid of the two 
pair-connectedness functions, the scaling analysis in the Appendix~\ref{appdc} reveals 
that asymptotically 
\begin{equation}
	2/\tilde{z}_a = z_a = 2/\tilde{z}_{b} ,
	\label{eqs:zexps}
\end{equation}
see Eq.~\eqref{eqs:a14a}.
This argument explains why the exponents $2/\tilde{z}_a$ and $2/\tilde{z}_b$ are
measured to be equal within numerical errors, and suggests their interpretation as the
ultimate dynamical critical exponent $z = z_a$ for the (C)PCPD.
Other subdominant diverging length and time scales then generate corrections to scaling 
in intermediate, but potentially long-lasting stages; see Eq.~\eqref{eqs:a15} for a 
specific example. 
The detailed competition between the various length and/or time scales of course 
depends on the measured quantity and can thus give rise to rather intricate and 
confusing scenarios in numerical simulations.

Furthermore, this multiple-scale picture based on the prevalence of two interacting
species may also provide a plausible explanation for the observed apparent dependences
of certain critical exponents on the diffusion rate in the original PCPD
\cite{odor2000critical,odor2003criti,dickman2002nonun,de2006moment}.
Taking the dynamical exponent $z$ in the literature as an example, one notices that $z$ 
tended to assume larger values for smaller diffusion rates
 \cite{henkel2004non, smallenburg2008univ}, and their values were measured closer to 
$2$ when determined via the finite-size scaling for the mean extinction time 
$\langle\tau\rangle = L^{z}$ 
\cite{carlon2001crit, odor2003criti, dickman2002nonun, de2006moment}; in contrast,
when $z$ was estimated through $R(t)\sim t^{1/z}$, the data resulted in markedly 
smaller values \cite{hinrichsen2001pair, noh2004univer, kwon2007double}. 
We propose that $z$ will ultimately reach $z_a \approx 1.65$, yet possibly after a 
rather long crossover time; for small diffusivities, this extended crossover regime may 
indeed extended to very long time periods. 

On the one hand, most PCPD implementations with small diffusivities are setup in the 
more reactive range, wherein the local processes \eqref{eqs16b} and \eqref{eqs16c} 
happen frequently, leaving the slowly diffusing solitary particles $B$ to occupy large 
regions in the system, until finally a large spanning cluster dominates the system's 
features.
Hence all (auto)correlations are affected to a large extent by the pure diffusive dynamics 
of the $B$ particles, and measurements of $z$  (via $2/\tilde{z}_b$ and 
$2/\tilde{z}_1$) will give results shifted closer to the purely diffusive value $2$.
On the other hand, the scaling relation $\langle\tau\rangle\sim L^{z}$ usually probes a 
shorter time scale than the mean-square displacement measurements 
$R(t)^2 \sim t^{2/z}$, because the latter is averaged over survival runs and the 
accessible time scale is in principle not limited by the system size if dynamically generated 
lists are used \cite{hinrichsen2006phase}. 
Hence the former method can more easily lead to a larger observed $z$ value,
especially for smaller diffusion rates. 
It is also worth noticing that our analysis for the dynamical exponents is consistent with 
a double domain structure analysis, c.f. Ref.~\cite{kwon2007double} (and 
Table~ \ref{tab:critexp}), with the correspondences 
$z_a=2/\tilde{z}_a \leftrightarrow Z_p$ and $2/\tilde{z}_b \leftrightarrow Z$ (not 
$Z_U$ for the uncoupled region in the paper) for the dynamical exponents of the coupled 
region and the whole domain.
$Z$ then precisely corresponds to the dynamical exponent $z$ in the literature. 
As asserted in Ref.~\cite{kwon2007double}, $Z\leftrightarrow 2/\tilde{z}_b$ 
asymptotically crosses over to $Z_p\leftrightarrow z_a$.

Finally, the hyperscaling relations \eqref{eqs:a14b} and \eqref{eqs:a14c} relate the 
initial slip exponents $\Theta_a$ and $\Theta_b$ to the decay exponents $\delta_a$ and 
$\delta_b$ in a natural manner. 
Since in one dimension, our CPCPD seed simulation data indicate that 
$\delta_a \simeq \delta_b$ (and $\delta_2 \simeq \delta_1$), it is not surprising to also 
find $\Theta_a \simeq \Theta_b$ ($\Theta_2 \simeq \Theta_1$). 
To summarize this subsection, the CPCPD provides an illuminating new perspective 
to decipher the hitherto controversial scaling features of the PCPD in terms of multiple
competing length and time scales.

\subsection{Moment ratios \label{sec3.4}} 

In addition to critical exponents, moment ratios have also been demonstrated to be an 
important tool for classifying universality classes \cite{dickman1998moment}. 
The $n$th order parameter moments
\begin{equation}
	m_n(t) = \left\langle \rho_s^n(t) \right\rangle 
\end{equation}
are measured at the critical point of a finite system with linear size $L$ after sufficient 
time $t$ has elapsed for it to have reached the quasi-stationary regime (hence the 
label `$s$') \cite{dickman2002quasi}. 
Moment ratios of the form $m_n / (m_r^i m_s^j)$ then become universal in the limit 
$L \to \infty$, provided $n = ri + sj$. 
In addition, the ratio $K_4 / K_2^2$ involving the cumulants 
\begin{equation}
	K_2 = m_2 - m_1^2
\end{equation}
and
\begin{equation}
	\quad K_4 = m_4 - 4 m_3 m_1 - 3 m_2^2 + 12 m_2 m_1^2 - 6 m_1^4
\end{equation}
is universal as well \cite{dickman1998moment}. 

In many ``conventional'' models, the transit to the quasi-stationary regime occurs after a 
relatively short period of time ($t \ll L^{z}$), and these moment ratios converge
correspondingly quickly with respect to $L$. 
However, due to the long crossover time to reach its asymptotic scaling regime, the 
PCPD single-particle moment ratios anomalously display apparently unlimited growth 
with increasing system size even if only the ``reactive sector'', with processes restricted 
to the subspace with at least one particle pair present, is considered 
\cite{dickman2002nonun}. 
In contrast, moment ratio convergences for consecutive pairs were only observed after
exceedingly long transient time ($\sim 10^9$ MCS) and for rather large systems 
($L=40,960$); in addition, the moment ratios for single particles continue to grow with 
$L$ but seem to coincide with those of consecutive pairs only at much larger system 
sizes \cite{de2006moment}.

\begin{figure}[!ptb]
 \begin{center}
  \includegraphics[width=0.48\textwidth]{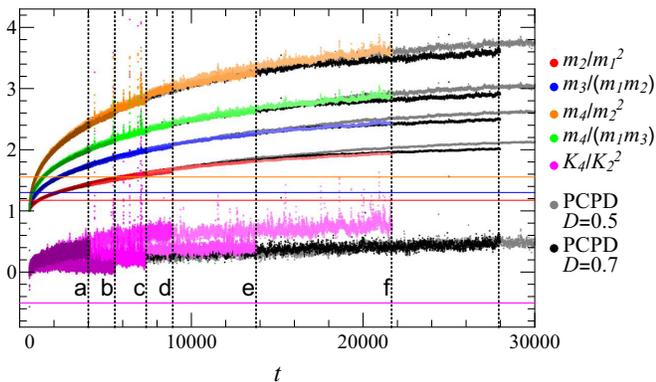}
 \end{center}
 \caption{Temporal evolution curves for the moment ratios of the CPCPD and the PCPD;
	 the gray and black lines show the reference
	 data for the PCPD (gray: $L=300$, $T=30,000$, $D=0.5$, and $\mu_c=0.13353$;
	 black: $L=300$, $T=30,000$, $D=0.7$, and $\mu_c=0.15746$).
	 The colorful graphs display CPCPD results obtained for different system 
	 sizes $L$, simulation time $T$, and diffusion rates $D_A$ and
	$D_B$ that have been rescaled 
	with respect to the time axis in order to achieve collapse with the reference curves.
	The dashed vertical lines (a-f) indicate at what rescaled times $t$ the simulation
	processes were terminated; 
 	a:~$L=1,000$, $T=500,000$, $D_A=0.01$, $D_B=0.02$, 
	$\mu_c=0.03539$; 
	b:~$L=1,000$, $T=200,000$, $D_A=0.2$, $D_B=0.22$ , 
	$\mu_c=0.08882$; 
	c:~$L=1,000$, $T=200,000$, $D_A=0.3$, $D_B=0.33$,
	$\mu_c=0.11054$;
	d:~$L=300$, $T=90,000$, $D_A=0.01$, $D_B=0.02$,
	$\mu_c=0.03539$; 
	e:~$L=300$, $T=50,000$, $D_A=0.2$, $D_B=0.22$, $\mu_c=0.08882$;
       f:~$L=300$, $T=50,000$, $D_A=0.3$, $D_B=0.33$, $\mu_c=0.11054$.
	All simulations were started from a homogeneous configuration with sites occupied 	
	by $B$ particles; the results were averaged over $50,000$ independent runs. 
	Owing to the smaller system sizes $L$, the accessible simulation durations $T$ were
	relatively short. The critical points were estimated on systems with 	$L=10,000$, and 
      their slight shifts for smaller systems due to finite-size effectes were omitted. 
      The horizontal lines mark the DP reference values \cite{dickman1998moment}.}
 \label{fig:moment}
\end{figure}
Henceforth, we focus on comparing the moment ratios of PCPD consecutive pairs and of
CPCPD pair particles $A$.
Yet instead of elaborately computing their asymptotic universal values, we report a novel 
method that compares the temporally evolving moment ratio trends, to demonstrate that
the two models attain the same crossover behaviors after the time axis is appropriately 
rescaled. 
To this end, we measure the moments $m_n(t) = \langle a(t)^n \rangle$ and 
$\langle \rho_2(t)^n \rangle$, respectively, and their ratios for all time steps.
The averages are performed over surviving runs, which can be conveniently achieved by 
the ``quasi-stationary'' simulation method \cite{de2005simulate,de2006moment}, even
though in this study it is not necessary for the quasi-stationary regime to be reached. 
In our implementation of this method, we accumulate a list of $2,000$ previously visited
surviving configurations.
During each Monte Carlo step, one of the randomly picked states among those in the 
repository will be substituted with the visited surviving configuration with a small 
probability $0.005$. 
Then, whenever an extinction event is imminent, the system replaces its current 
configuration with another one that is randomly selected from this list. 
Furthermore, since we are only interested in the temporal behaviors of the CPDPD and 
the PCPD, we may simply compare the moment ratios of the two models obtained from
their entire model spaces, rather than focusing on the reactive subspace as in 
Ref.~\cite{de2006moment}.

Figure~\ref{fig:moment} shows the time evolution tracks of the moment ratios for both
the CPCPD and the  PCPD, where the PCPD curves for system size $L=300$ and 
diffusivity $D=0.7$ as well as the PCPD curves for $L=300$ and $D=0.5$ have been 
colored black and gray respectively for reference. 
All other curves corresponding to different CPCPD systems with different $L$, 
simulation durations $T$, and diffusion rates were then rescaled with respect to the time
axis, so that they all collapse to the corresponding PCPD reference graphs for the larger 
diffusivity $D=0.7$.
Except for the data for the moment ratio $K_4(t)/K_2(t)^2$ (bottom) which turn out
quite noisy, all other curves display a quite satisfactory collapse.
We note that starting from common initial values, the moment ratios for two distinct 
realizations of the same universality class will of course assume all values between the 
initial and eventual universal ones, but need not at all display identical crossover features.
The remarkable complete data collapse we observe in Fig.~\ref{fig:moment} hence
provides strong numerical evidence that, on the one hand, PCPD systems with
large enough diffusivities essentially evolve with the same dynamics; on the other hand, 
regardless of system sizes and diffusion rates, the CPCPD displays the very same 
stochastic dynamics as the PCPD, albeit with an overall slower pace, as becomes apparent 
upon comparison of the two models run on lattices with the same size.
We attribute this slower kinetics for the CPCPD to the required intermediate pair 
production processes that occur at a finite rate $\tau$. 
 
Specifically, even the moment ratio data for the CPCPD with fairly small diffusion rates
($D_A=0.01$ and $D_B=0.02$, c.f.~the ``a'' and ``d'' lines in Fig.~\ref{fig:moment}) 
collapse very well with those of the PCPD, hinting that the CPCPD is less afflicted by the 
strong corrections to scaling as compared to the PCPD for small diffusivities $D$
\cite{park2014critical}.
(The PCPD results with $D=0.1$ in fact do not display good data collapse with the other
curves; data not shown.)
It should be noted that the intrinsic slow crossover features in both the CPCPD and PCPD 
are already encoded in the moment ratio curves. 
Taking the reference black curves in Fig.~\ref{fig:moment} as an example, it is apparent
that the moment ratios still show no signs of stationarity even for $t \sim L^z$, but their 
slopes flatten for increasing $t$, implying that even if these moment ratios reach some 
universal asymptotic values, these would exceed the results inferred solely from the
reactive subspace data as in Ref.~\cite{de2006moment}. 
Nonetheless, given that the moment ratios for single particles are expected to coincide 
with these of particle pairs (consecutive particles) at very large system sizes 
\cite{de2006moment}, and that the moment ratios measured in Fig.~\ref{fig:moment} 
at larger times are clearly very different from the established values for both the DP 
(see the horizontal lines in Fig.~\ref{fig:moment}) and the PC universality classes 
\cite{dickman1998moment},  we believe it is safe to conclude that the critical PCPD (and
CPCPD) belongs to neither of these two universality classes.

\section{Summary \label{sec4}}

In this work, we have introduced a coupled two-species model  (CPCPD) to represent the 
pair contact process with diffusion (PCPD). 
Introducing intermediate pair production processes effectively ``fine-grains'' the PCPD to
adequately capture its internal noise induced by the stochastic nonlinear reactions. 
Similar species separations may also be applicable for other higher-order processes, such
as the triplet-contact process \cite{park2002phase, kockelkoren2003abs}. 
The analysis of the associated coupled mean-field rate equations suggests that, as the 
PCPD, the CPCPD also displays algebraic density decays both in the inactive phase and at 
criticality. 
Unlike conventional models where continuous phase transitions occur when linear 
``mass'' terms vanish, the CPCPD critical point is driven by the balance between the 
density changes of the competing single-particle and particle-pair species. 

Our extensive Monte Carlo simulations demonstrate that the scaling properties of the 
CPCPD are fully consistent with those of the PCPD. 
In the inactive phase, both models are governed by the power laws of the pure 
diffusion-limited pair annihilation process. 
At criticality, the decay exponents of the two models agree with each other in one, two, 
and three dimensions, with the mean-field exponents recovered in dimensions 
$d > 2$, indicating that the upper critical dimension of the CPCPD and PCPD is 
$d_c = 2$.
Due to a superposition of the as yet unknown logarithmic fluctuation corrections at 
$d_c$,  as well as the sizeable intrinsic corrections to scaling in these models, the 
effective critical decay exponents for the single-particle densities of both models show 
deviations from the mean-field value $1/2$ in two dimensions. 
The power law decay for the pair or consecutive pair order parameters may also be 
affected by these various scaling corrections at the critical dimension, but our data 
indicate that these deviations are much weaker for the pair observables.
This allows us to draw more consistent conclusions for our two- and three-dimensional 
simulations and their comparison with the mean-field predictions than in 
Ref.~\cite{odor2002phase}. 

To properly interpret the seed simulation data, we argue that two distinct 
pair-connectedness functions should be considered, and therefore also two sets of seed
simulation exponents for each order parameter must be introduced. 
In this manner, the conceptual difficulty residing in the original PCPD concerning seed
simulations is resolved in a natural way. 
Owing to its formulation in terms of two particle species, the CPCPD inevitably involves 
several length and associated time scales describing the correlations between the species. 
We propose that the corrections to scaling in both the CPCPD and PCPD can be ascribed 
to the competition between these length and/or time scales. 
Positing that in the asymptotic large-scale and long-time regime, one of these scales
will become dominant, scaling analysis leads to hyperscaling relations for the two sets of
seed simulation exponents; moreover, the system is ultimately governed by the 
dynamical exponent $z_a$ for the particle \textit{pairs}.
As for the detailed mechanism of how these multiple length and time scales may drive
corrections to scaling, we advocate for further quantitative studies on the two involved
pair-connectedness functions and the (cross-)correlation lengths / times of the CPCPD.

We have found that after straightforward time rescaling, the moment ratios for the 
CPCPD pair particles coincide with those of consecutive pairs in the PCPD in their full 
temporal crossover behavior.
Hence we conclude that the dynamical behavior of both models is essentially identical,
only that the CPCPD evolves with a slower kinetics. 
Consequently we interpret the CPCPD and PCPD to be truly equivalent, not merely with
respect to their asymptotic universal scaling properties, but throughout their time
evolution.
Finally, we point out that the asymptotic values of the moment ratios of the (C)PCPD
appear to differ markedly from those of the DP and the PC universality classes 
\cite{dickman1998moment}.
Since the possibility that the PCPD belongs to the PC equivalence class has long been 
ruled out \cite{henkel2004non}, we (again) point out two major and fundamental 
distinctions that separate the (C)PCPD from the generic directed percolation university 
class:
First, their upper critical dimensions are different, $d_c=4$ for DP, whereas $d_c=2$ 
forthe (C)PCPD.
Second, the generic DP inactive phase is normally characterized by exponential 
decay of the particle density as well as spatial and temporal correlations, with a 
notable exception reported for certain higher-order processes such as $3B \to 4B$ and
$2B \to \emptyset$, as studied in, e.g.,~Ref.~\cite{kockelkoren2003abs}. 
In stark contrast, the (C)PCPD inactive phase is exclusively governed by the 
scale-invariant power laws of diffusion-limited pair annihilation kinetics. 
Since the CPCPD comprises only at most binary reactions (and reaction products), 
and explicitly accounts for the crucial particle pairs as an independent species, we hope 
that future field-theoretic studies may elucidate the precise nature of its critical 
properties, and hence those of the PCPD.

\begin{acknowledgments}
We are indebted to Geza \'{O}dor for insightful discussions and suggestions.
SD acknowledges a fellowship from the China Scholarship Council (CSC) under Grant 
CSC No.~201806770029.
WL acknowledges support from the National Science Foundation of China (NSFC) through
research grant No.~61873104 and the Programme of Introducing Talents of Discipline to
Universities under Grant No.~B08033.
UCT's research was sponsored by the Army Research Office and was accomplished under 
Grant No.~W911NF-17-1-0156. 
The views and conclusions contained in this document are those of the authors and 
should not be interpreted as representing the official policies, either expressed or 
implied, of the Army Research Office or the U.S. Government. 
The U.S. Government is authorized to reproduce and distribute reprints for 
Government purposes notwithstanding any copyright notation herein. 
\end{acknowledgments}

\appendix

\section{Mean-field effective density decay exponents \label{appda}}
\begin{figure}[!hptb]
        \begin{center}
                \begin{tabular}{c}
                        \includegraphics[width=0.4\textwidth]{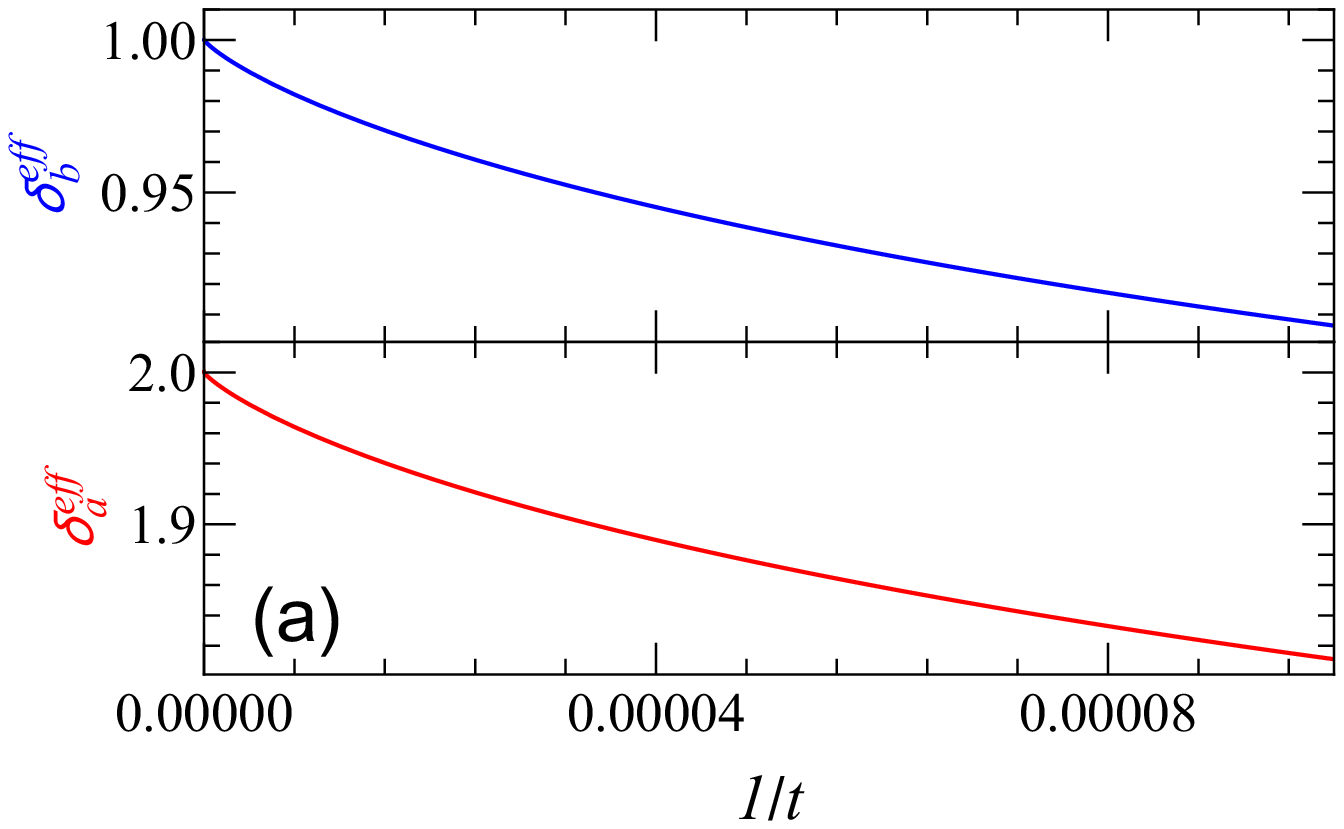}\\
                        \includegraphics[width=0.4\textwidth]{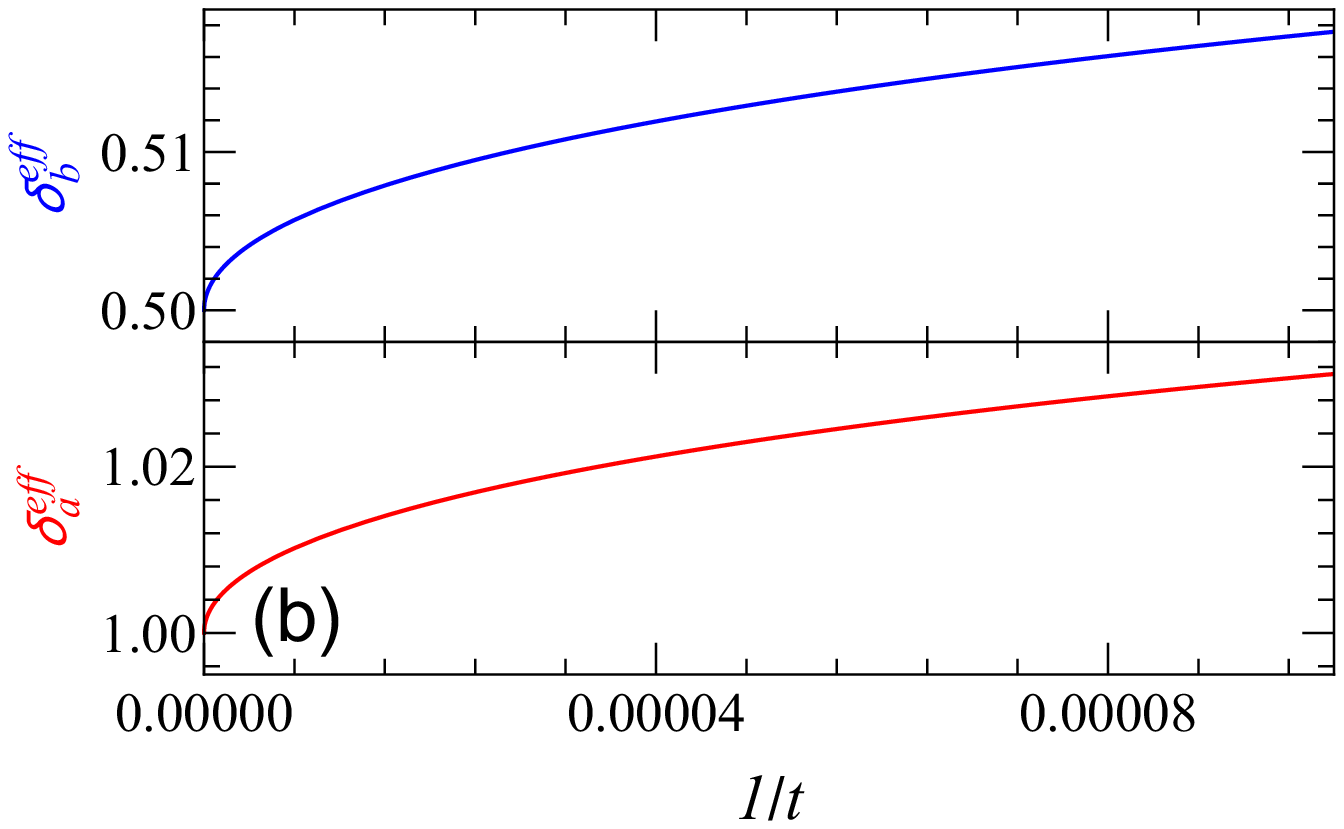}
                \end{tabular}
        \end{center}
        \caption{Mean-field effective density decay exponents for (a) the
	inactive-phase curves and (b) the critical curves in Fig.~\ref{fig:decay}. 
	The displayed time scale ranges from $10^4$ to 	$10^{10}$.}
        \label{fig:mfeffexp}
\end{figure}

The slow clossover behavior of the (C)PCPD is even manifested in the mean-field
regime. Figure \ref{fig:mfeffexp} shows the local slopes [Eq.~\eqref{eqs:eff}]
of the density decay exponents of the CPCPD in the inactive phase [(a)] and at 
criticality [(b)]. Therefore even the critical point in the mean-field regime
may not be easily fixed by applying standard methods and one usually expects to 
observe exponent values slightly larger than the mean-field predictions within
a relatively finite time scale, as observed in all critical three-dimensional 
cases in this paper.

\section{Critical density decay exponents for the PCPD with parity conservation 
	\label{appdb}}
\begin{figure}[!htpb]
	\begin{center}
		\includegraphics[width=0.42\textwidth]{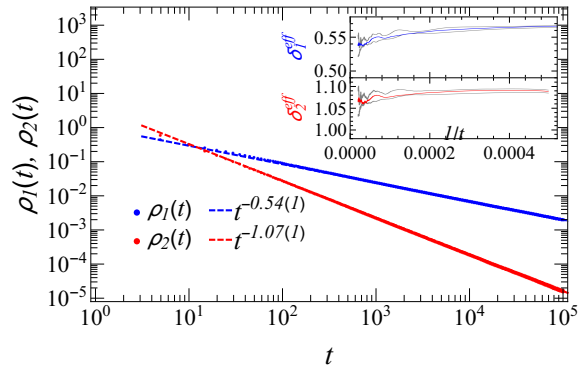}
	\end{center}
	\caption{Critical density decay results for the parity-conserving PCPD in three 
	dimensions with $L=320$ and $\mu=0.4868$, averaged over $10$ runs; in 
	the inset, the curves pertain to $\mu=0.4866, 0.4868$, and $0.4870$. 
	All other setups are the same as described in the 	caption of Fig.~\ref{fig:crit}.}
	\label{fig:3dpar}
\end{figure}

In this section, we examine the effects of particle number parity conservation 
on the critical density decay exponents of the PCPD. 
We again apply the simulation method in Sec.~\ref{sec3.1} and, in accord with 
Ref.~\cite{odor2002phase}, for the branching process $2B \to 4B$, two new 
particles are evenly generated at the available neighboring sites of a selected pair. 
In one dimension, parity conservation makes no difference (data not shown) as 
suggested in the literature (see, e.g., Ref.~\cite{park2001binary}). 
In three dimensions, as shown in Fig.~\ref{fig:3dpar}, the mean-field CPCPD 
density decay exponents are recovered.

\begin{figure}[!htpb]
	\begin{center}
	\begin{tabular}{c}
		\includegraphics[width=0.42\textwidth]{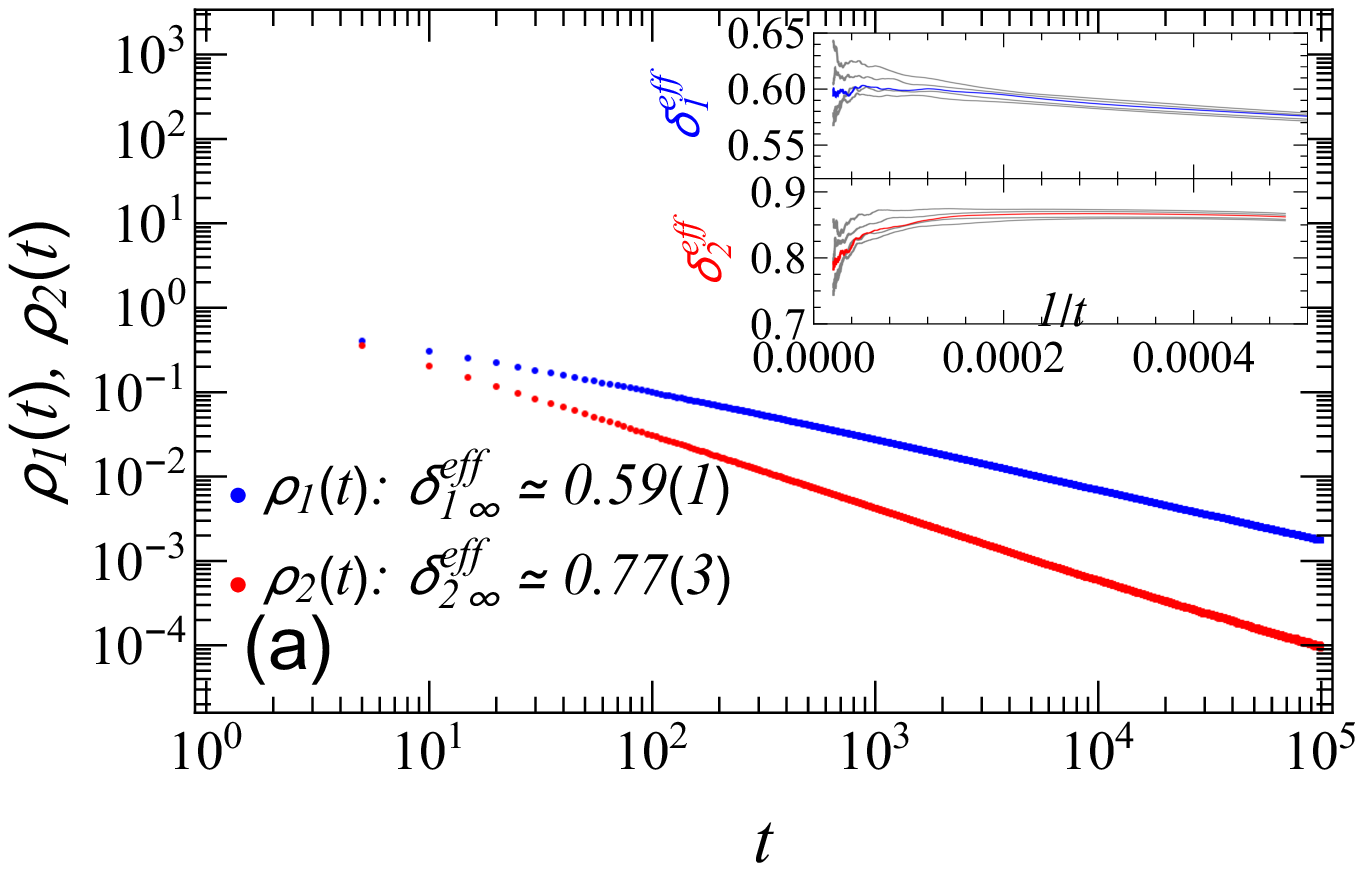}\\
		\includegraphics[width=0.42\textwidth]{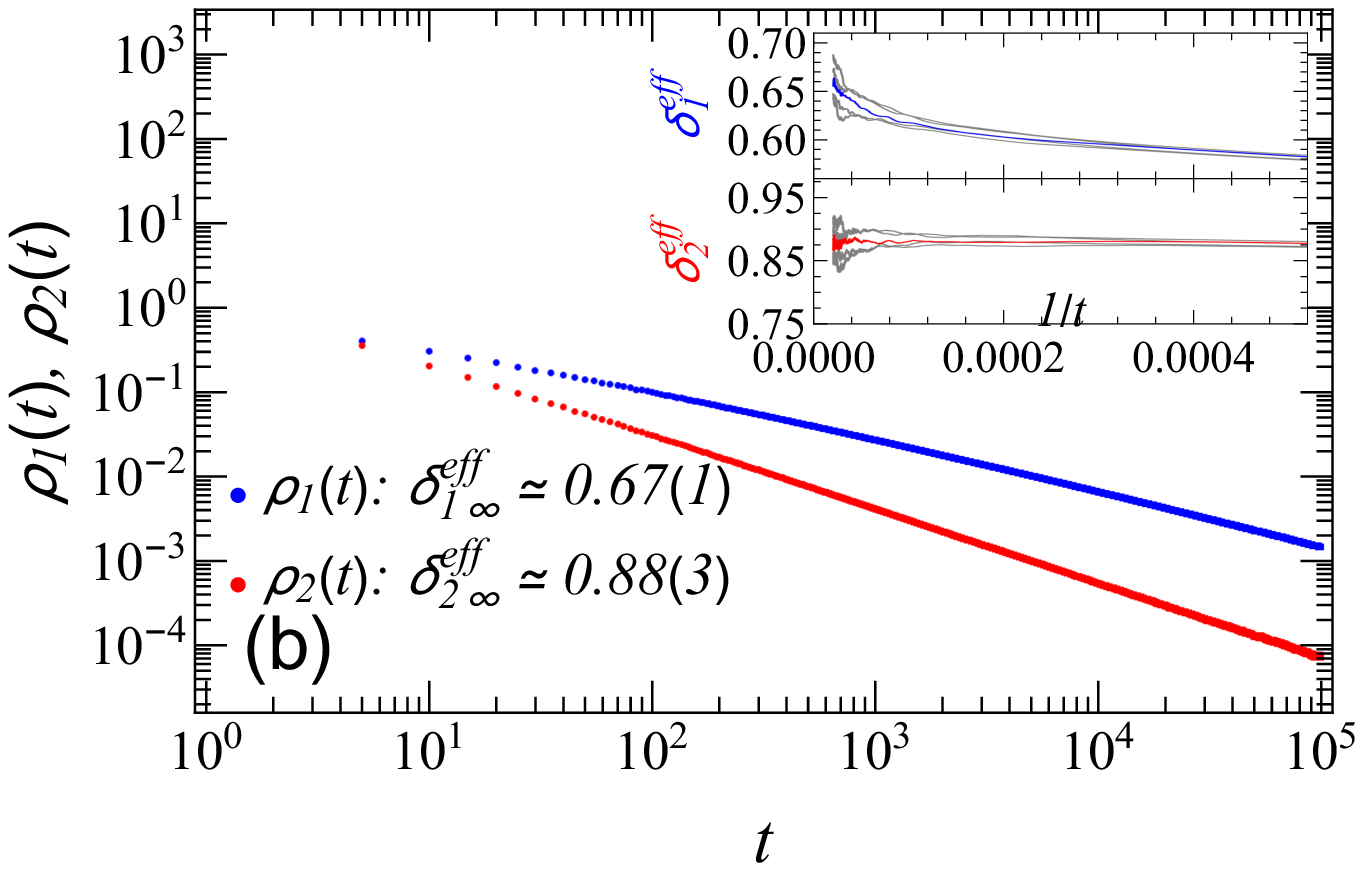}
	\end{tabular}
	\end{center}
	\caption{Critical density decay results for the parity-conserving PCPD in two 
	dimensions with $L=640$, averaged over $1,000$ runs.
	The critical point is estimated according to (a) the asymptotic stationary value 
	for $\delta_{1}^{\mathrm{eff}}$ at $\mu=0.37850(5)$ (in the inset, from 
	bottom to top, the curves pertain to 
	$\mu=0.3783, 0.3784, 0.3785, 0.3786, 0.3787$) and (b) the asymptotic
	stationary value for $\delta_{2}^{\mathrm{eff}}$ at $\mu=0.37890(5)$ (in 
	the inset, from bottom to top, $\mu=0.3787, 0.3788, 0.3789, 0.3790, 0.3791$).
      All other setups are the same as described in the caption of Fig.~\ref{fig:crit}.}
	\label{fig:2dpar}
\end{figure}

\begin{figure}[!htpb]
	\begin{center}
		\includegraphics[width=0.42\textwidth]{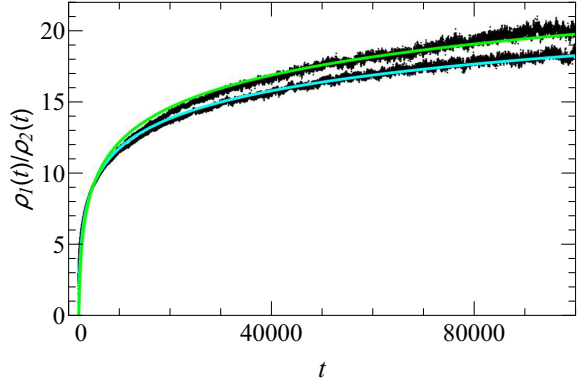}
	\end{center}
	\caption{$\rho_1(t)/\rho_2(t)$ and logarithmic fit for the data
		from Fig.~\ref{fig:2dpar} (a) (bottom) and
		Fig.~\ref{fig:2dpar} (b) (top).}
	\label{fig:rhoratio}
\end{figure}
In two dimensions, however, particle number parity conservation indeed plays 
a significant role. 
Due to the superposition of various unknown corrections to scaling, such as 
logarithmic corrections and the slow crossover behavior as demonstrated in 
App.~\ref{appda}, it is not trivial to fix the critical point without explicitly taking 
into account these corrections. 
Omitting the complications caused by these corrections first, one may estimate the 
critical point either according to the stationary effective exponent value of the particle 
density [Fig.~\ref{fig:2dpar} (a)] or according to that of the pair density 
[Fig.~\ref{fig:2dpar} (b)]. 
As shown in Fig.~\ref{fig:2dpar} (a) and (b), respectively, we neither obtained 
$\delta_{1}^{\mathrm{eff}}\approx 0.5$ as in Ref.~\cite{odor2002phase}, nor did 
we find $\delta_{2}^{\mathrm{eff}} \approx 1$, in contrast to the results in
Fig.~\ref{fig:crit} (c) and (d) for the standard (C)PCPD. 
We also checked another version of the parity-conserving PCPD where newly created
particles are generated along the chosen direction in two and three dimensions, the 
results are the same as the above within error margins (data not shown).
Furthermore, we indeed observed the scaling $\rho_1(t)/\rho_2(t) \sim \ln(t)$ (see 
Fig.~\ref{fig:rhoratio}) as in Refs.~\cite{odor2002phase,park2005driven}. 
Hence we believe that the confusing situation in two dimensions is really due to the 
unknown strong corrections to scaling, and the discrepancies between our results in 
this paper (with or without parity conservation) and the results in 
Ref.~\cite{odor2002phase} might be caused by these large corrections and our 
different microscopic implementations.

\section{Pair-connectedness functions and seed simulation exponents \label{appdc}}

The dynamical scaling behavior of conventional one-species systems starting from 
non-trivial configurations can be derived from the pair-connectedness function
$\Upsilon(t_1,t_2;\mathbf{r}_1,\mathbf{r}_2;\Delta)$ that probes the causal 
connection between two space-time points \cite{henkel2008non}. 
It is defined as the probability of a cluster generated at seeding point 
$(t_1, \mathbf{r}_1)$ activates site $\mathbf{r}_2$ at time $t_2$.
Translational invariance in space and time allows us to write the pair-connectedness 
function as $\Upsilon(t,r;\tau)$, where $r = |\mathbf{r}_2 - \mathbf{r}_1|$ and 
$t = t_2 - t_1 \ge 0$. 
In the CPCPD, even though there are two particle species, a cluster can only be
generated from an $A$ seed; therefore we extend the above definition to include two 
pair-connectedness functions, $\Upsilon_{aa}(t, \mathbf{r}; \Delta)$ and 
$\Upsilon_{ab}(t, \mathbf{r}; \Delta)$, which represent the probabilities of finding an
$A$ or $B$ particle at space-time point $(t,\mathbf{r})$, in a sequence of stochastic
processes starting from an $A$ seed located at $\mathbf{r}=0$ at time $t=0$. 

Henceforth we follow the derivations in Sec.~4.4.2 and thereafter of 
Ref.~\cite{henkel2008non} to analyze the ensuing critical scaling of the two
pair-connectedness functions. 
In the steady-state limit $t \to \infty$, the pair-connectedness functions on the one hand
depend on the probability $P_{\mathrm{inf}}(\Delta) \sim \Delta^{\beta'}$ that the
seed at $(t=0, \mathbf{r}=0)$ generates an infinite cluster, and on the other hand, the
probability that the chosen site $\mathbf{r}$ belongs to the infinite cluster, which is just
the steady-state density $a_s(\Delta)$ or $b_s(\Delta)$, respectively, whence
\begin{IEEEeqnarray}{rCl}
	\label{eqs:a1}
	\lim_{t\to\infty}\Upsilon_{aa}(t,r;\Delta) &=& a_s(\Delta)
	P_{\mathrm{inf}}(\Delta) \sim \Delta^{\beta_a+\beta'}\,, \qquad
	\IEEEyesnumber* \IEEEyessubnumber* \label{eqs:a1a}\\
        \lim_{t\to\infty}\Upsilon_{ab}(t,r;\Delta) &=& b_s(\Delta)
	P_{\mathrm{inf}}(\Delta) \sim \Delta^{\beta_b+\beta'}\,. \qquad
	\label{eqs:a1b}
\end{IEEEeqnarray}
Therefore the pair-connectedness functions should scale as
\begin{IEEEeqnarray}{rCl}
	\label{eqs:a2}
	\Upsilon_{aa}&\mapsto & \lambda^{\beta_a+\beta'}\Upsilon_{aa}\,,
	\IEEEyesnumber* \IEEEyessubnumber* \label{eqs:a2a} \\
        \Upsilon_{ab}&\mapsto & \lambda^{\beta_b+\beta'}\Upsilon_{ab}\,,
\end{IEEEeqnarray}
upon rescaling $\Delta \to \lambda \Delta$. 
The (C)PCPD system involves several length and time scales related to the 
(cross-)correlations between identical species or different species (see the discussion in 
Sec.~\ref{sec3.3}). 
When the system asymptotically becomes scale-invariant in the limit $t\to\infty$, (only) 
one of the correlation length (time) scales dominates the long-distance (-time) features. 
Without any prior knowledge about the relative relations of these length (time) scales,
we assume it to be
$\xi_{X} \sim \Delta^{-\nu_{X}}$ ($t_{c X} \sim \Delta^{-z_{X}\nu_{X}}$), where the 
label ``$X$'' is to be determined later after the meanings of the exponents 
$2/\tilde{z}_a$ and $2/\tilde{z}_b$ become clear. 
Next dimensionless scaling functions can be constructed with independent arguments  
$t / t_{c X}$, $x / \xi_{X}$, and $\Delta$. 
The two pair-connectedness functions then take the following scaling forms
\begin{IEEEeqnarray}{rCl}
	\label{eqs:a3}
	\Upsilon_{aa}(t,r;\Delta)\simeq \lambda^{-\beta_a-\beta'}
	\tilde{\Upsilon}_{aa}(\lambda^{-z_{X}\nu_{X}} t,
	\lambda^{-\nu_{X}}r; \lambda \Delta)\, , \quad \qquad \IEEEyesnumber*
	\IEEEyessubnumber* \label{eqs:a3a} \\
        \Upsilon_{ab}(t,r;\Delta)\simeq \lambda^{-\beta_b-\beta'}
	\tilde{\Upsilon}_{ab}(\lambda^{-z_{X}\nu_{X}} t,
	\lambda^{-\nu_{X}}r; \lambda \Delta)\, . \quad \qquad \label{eqs:a3b}
\end{IEEEeqnarray}

Furthermore, sufficiently close to criticality, the quantities studied in seed simulations 
follow the scaling forms
\begin{widetext}
\begin{IEEEeqnarray}{rCl}
	\label{eqs:a4}
	P_{\mathrm{sur}}(t; \Delta) \simeq  \lambda^{-\delta'z_{X}\nu_{X}}
	\tilde{P}_{\mathrm{sur}}(\lambda^{-z_{X}\nu_{X}} t; \lambda
	\Delta)\,, &\quad & \IEEEyesnumber* \IEEEyessubnumber*
	\label{eqs:a4a} \\
	N_{a}(t; \Delta)  \simeq  \lambda^{\Theta_a z_{X}\nu_{X}}
	\tilde{N}_{a}(\lambda^{-z_{X}\nu_{X}} t; \lambda \Delta)\,,
	& \quad & N_{b}(t; \Delta)  \simeq \lambda^{\Theta_b z_{X}\nu_{X}}
	\tilde{N}_{b}(\lambda^{-z_{X}\nu_{X}} t; \lambda
	\Delta)\,,\qquad \label{eqs:a4b}\\
	R_{a}(t ; \Delta) \simeq \lambda^{\nu_{aX}} \tilde{R}_{a} (
	\lambda^{-z_{X}\nu_{X}} t; \lambda \Delta)\,, & \quad &
        R_{b}(t ; \Delta) \simeq \lambda^{\nu_{bX}} \tilde{R}_{b} (
	\lambda^{-z_{X}\nu_{X}} t; \lambda \Delta) \, , \label{eqs:a4c}
\end{IEEEeqnarray}
\end{widetext}
where in the last line we have taken caution not to set $\nu_{aX}=\nu_{X}=\nu_{bX}$, 
because the dynamical exponents for $R_a$ and $R_b$ may be different from each 
other and $z_{X}$.

The exponent $\Theta_a$ can be obtained by expressing the average number of 
particles $N_a(t,\Delta)$ in terms of the pair-connectedness function via
\begin{IEEEeqnarray}{rCl}
	N_{a}(t ; \Delta) &=& \int \mathrm{d} \mathbf{r}
	\Upsilon_{aa}(t,|\mathbf{r}|; \Delta) \nonumber \\
        &=& \int \mathrm{d} \mathbf{r} \lambda^{-\beta_a-\beta'}
	\tilde{\Upsilon}_{aa}(\lambda^{-z_{X}\nu_{X}} t,
	\lambda^{-\nu_{X}} r ; \lambda \Delta) \nonumber \\
        &\simeq & \lambda^{d \nu_{X}-\beta_a-\beta'} \int \mathrm{d}
	\mathbf{r} \tilde{\Upsilon}_{ab} (\lambda^{-z_{X}\nu_{X}}t, r ;
	\lambda \Delta) \nonumber \\
	&\simeq & \lambda^{d\nu_{X}-\beta_a-\beta'}
	\tilde{N}_a(\lambda^{-z_{X}\nu_{X}}t; \lambda \Delta) \, ,
	\label{eqs:a5}
\end{IEEEeqnarray}
where we have inserted Eq.~\eqref{eqs:a3a} and substituted 
$r \to \lambda^{\nu_{X}} r$ in the integral. 
Comparison with Eq.~\eqref{eqs:a4b} yields the hyperscaling relation
\begin{equation}
	\frac{d}{z_{X}} = \Theta_a+\frac{\beta_a}{z_{X}
	\nu_{X}}+\frac{\beta'}{z_{X}\nu_{X}} \, .
	\label{eqs:a6}
\end{equation}
Similarly for $\Theta_b$, by exploiting the pair-connectedness function 
$\Upsilon_{ab}$ and Eq.~\eqref{eqs:a3b}, and comparing the final expression
to Eq.~\eqref{eqs:a4b}, we obtain
\begin{equation}
	\frac{d}{z_{X}} = \Theta_b+\frac{\beta_b}{z_{X}
	\nu_{X}}+\frac{\beta'}{z_{X}\nu_{X}} \, .
	\label{eqs:a7}
\end{equation}

The last two terms in Eq.~\eqref{eqs:a6} and Eq.~\eqref{eqs:a7} can be related to
other critical exponents. 
In the active phase, the ultimate survival probability is just the probability of being part
of the infinite cluster, $\lim_{t\to\infty} P_{\mathrm{sur}}(t,\Delta) =
P_{\mathrm{inf}}(\Delta) \sim \Delta^{\beta'}$, which implies
\begin{equation}
	\delta'=\frac{\beta'}{z_{X}\nu_{X}} \, .
	\label{eqs:a8}
\end{equation}
In addition, the densities for the two species obey the scaling forms
\begin{IEEEeqnarray}{rCl}
	\label{eqs:a9}
	a(t;\Delta)& \simeq & \lambda^{\beta_a}
	\tilde{a}(\lambda^{-z_{X}\nu_{X}}; \lambda\Delta) \sim
	t^{-\delta_a}\,,
	\IEEEyesnumber* \IEEEyessubnumber* \label{eqs:a9a} \\
      b(t;\Delta)& \simeq & \lambda^{\beta_b} 
	\tilde{b}(\lambda^{-z_{X}\nu_{X}}; \lambda\Delta) \sim
	t^{-\delta_b}\,, \label{eqs:a9b}
\end{IEEEeqnarray}
which lead to the relations
\begin{equation}
	\delta_a=\frac{\beta_a}{z_{X}\nu_{X}} \, , \quad 
	\delta_b =
	\frac{\beta_b}{z_{X}\nu_{X}} \, .
	\label{eqs:a10}
\end{equation}
Inserting Eqs.~\eqref{eqs:a8} and \eqref{eqs:a10} into 
Eqs.~\eqref{eqs:a6} and \eqref{eqs:a7}, we thus arrive at the hyperscaling 
relations
\begin{IEEEeqnarray}{rCl}
	\label{eqs:a11}
	\Theta_a &=& \frac{d}{z_{X}} - \delta_a -\delta' \, ,
	\IEEEyesnumber* \IEEEyessubnumber* \label{eqs:a11a} \\
      \Theta_b &=& \frac{d}{z_{X}} - \delta_b -\delta' \, .
	\label{eqs:a11b}
\end{IEEEeqnarray}
As the numerical results in one dimension show that $\delta_a \approx \delta_b$, it
should follow that $\Theta_a \approx \Theta_b$, as indeed observed in the simulation
data.

Finally, a straightforward computation yields the exponents $\tilde{z}_a$ and 
$\tilde{z}_b$. 
To obtain $\tilde{z}_a$, we express $R_a^2$ as
\begin{IEEEeqnarray}{rCl}
	&&R_a(t; \Delta)^2=\langle|\mathbf{r}|^2\rangle 
	= \frac{1}{N_a(t)} \int \mathrm{d} \mathbf{r} r^2
	\Upsilon_{aa}(t,|\mathbf{r}|; \Delta) \nonumber \\
	&&\ = \frac{1}{N_a(t)} \int \mathrm{d} \mathbf{r} r^2 
	\lambda^{-\beta_a-\beta'}
	\tilde{\Upsilon}_{aa}(\lambda^{-z_{X}\nu_{X}} t,
	\lambda^{-\nu_{X}} r ; \lambda \Delta) \nonumber \\
	&&\ \simeq \frac{1}{N_a(t)}\lambda^{(d+2)\nu_{X}-\beta_a-\beta'} 
	\int \mathrm{d}
	\mathbf{r} r^2 \tilde{\Upsilon}_{ab} (\lambda^{-z_{X}\nu_{X}}t, r ;
	\lambda \Delta) \nonumber \\
	&&\ \simeq \lambda^{2\nu_{X}}
	\tilde{R}_a(\lambda^{-z_{X}\nu_{X}}t; \lambda \Delta)^2 \sim t^{2/z_{X}} \, .
	\label{eqs:a12}
\end{IEEEeqnarray}
Together with a similar analysis for $R_b(t; \Delta)^2$, we obtain the relation
\begin{equation}
	\label{eqs:a13}
	\frac{2}{\tilde{z}_a} = z_{X} = \frac{2}{\tilde{z}_b} \, . 
\end{equation}
Since both the mean-field analysis and the numerical results in one dimension, yielding
$R_a(t) > R_b(t)$, suggest that the correlations between the $A$ pair particles 
ultimately dominate the system's critical properties, we infer that all ``$X$'' labels above 
are to be replaced by ``$a$'', whence at last
\begin{IEEEeqnarray}{rCl}
	\label{eqs:a14}
	\frac{2}{\tilde{z}_a} &=& z_{a} = \frac{2}{\tilde{z}_b} \, ,
	\IEEEyesnumber* \IEEEyessubnumber* \label{eqs:a14a}\\
	\Theta_a &=& \frac{d}{z_{a}} - \delta_a -\delta' \, ,
	\label{eqs:a14b} \\
        \Theta_b &=& \frac{d}{z_{a}} - \delta_b -\delta' \, .
	\label{eqs:a14c}
\end{IEEEeqnarray}
However, in the finite run times accessible to numerical simulations, both 
$2/\tilde{z}_a$ and $2/\tilde{z}_b$ may be affected by corrections to scaling and hence
apparently still deviate from the asymptotic value of $z_{a}$ during extended crossover
periods. 
Note that Eq.~\eqref{eqs:a10} then resolves the apparent inconsistency 
$\delta_b = \beta_b / z_b \nu_b = 1 \neq \frac12$ in the mean-field approximation,
if separate length scales are used for each species for their corresponding scaling 
relations.

If we take into account the subdominant length scale, say 
$\xi_Y\sim \Delta^{-\nu_Y}$ with $\nu_Y<\nu_a$, the ratio of the two length scales
$\Delta^{-\nu_{Y}}/\Delta^{-\nu_{a}}$ may then enter as an irrelevant scaling field 
$\kappa$ \cite{shao2016quantum} in the pair-connected functions so that 
$\Upsilon(t,r;\Delta,\kappa)\simeq \lambda^{-\beta_a-\beta'} 
	\tilde{\Upsilon}(\lambda^{-z_{X}\nu_{X}} t ,
	\lambda^{-\nu_{X}}r; \lambda \Delta,
	\lambda^{\nu_Y-\nu_a}\kappa)$.
Provided that the scaling function $\tilde{\Upsilon}$ is analytic with respect to its fourth
argument, proceeding similarly as above, one obtains corrections to the leading scaling
\cite{wegner1972cor}
\begin{IEEEeqnarray}{rCl}
	R_a(t;\Delta)^2 &\simeq &
	\lambda^{2\nu_a}(1+\tilde{R}_a' \kappa \lambda^{\nu_Y-\nu_a})
	\tilde{R}_a(\lambda^{-z_a \nu_a}t; \lambda \Delta)^2 \nonumber \\
	& \sim & t^{2/z_a}(1+ \tilde{R}_a' \kappa t^{\frac{\nu_Y-\nu_a}{z_a
	\nu_a}}), \label{eqs:a15}
\end{IEEEeqnarray}
where the factor $\tilde{R}_a'$ relates to the integrals of 
$\partial_\kappa\tilde{\Upsilon}_{aa}$. 
When the two diverging length scales are close to each other, $\nu_Y\approx \nu_a$, 
the factor in the brackets only crosses over to $1$ after an extremely long time, causing
strong corrections to the measurements for $z_a$, such as observed in, 
e.g., Ref.~\cite{kwon2007double}.

\bibliographystyle{apsrev4-1}

%
\end{document}